\documentclass[11pt,a4paper]{article}
\DeclareUnicodeCharacter{2212}{\ensuremath{-}}
\usepackage{jheppub,xcolor}
\usepackage[utf8]{inputenc}
\usepackage{textcomp}
\usepackage{newunicodechar}
\newunicodechar{γ}{\ensuremath{\gamma}}
\usepackage{blindtext}
\usepackage{graphicx}
\usepackage{physics}
\usepackage{amsmath}
\usepackage{amsfonts}
\usepackage{amsthm}
\usepackage{amssymb}
\usepackage{caption}
\usepackage{booktabs}
\usepackage{subcaption}
\usepackage{slashed}
\usepackage{bigints}
\usepackage{setspace}
\usepackage{array}
\usepackage{comment}
\usepackage{multirow}
\usepackage{makecell} %  Allows line breaks in table cells
\definecolor{mygreen}{rgb}{0, 0.5019607843137255, 0}
\definecolor{Red}{rgb}{1.,0.,0.}
\definecolor{Blue}{rgb}{0.,0.,1.}
\newcommand{\CF}{C_F}
\newcommand{\CA}{C_A}
\newcommand{\qbar}{\overline{Q}}

\newcommand{\tbar}{\Bar{t}}
\newcommand{\bbar}{\Bar{b}}

\def\beq{\begin{equation}}
\def\eeq{\end{equation}}
\def\bsp#1\esp{\begin{split}#1\end{split}}

\newcommand{\cO}{\mathcal{O}}
\newcommand{\MSbar}{$\overline{\textrm{MS}}$}
\allowdisplaybreaks

\title{Two-loop renormalisation of the quark fields in the presence of four-fermion SMEFT operators}

\author[a]{Claude Duhr,}
\author[b]{Giuseppe Ventura,}
\author[b, c]{Eleni Vryonidou}

\emailAdd{cduhr@uni-bonn.de} \emailAdd{giuseppe.ventura@manchester.ac.uk} \emailAdd{vryonidou.eleni@ucy.ac.cy}
\affiliation[a]{Bethe Center for Theoretical Physics, Universit\"{a}t Bonn, Wegelerstrasse 10, D-53115, Germany}
\affiliation[b]{Department of Physics and Astronomy, University of Manchester, Oxford Road, Manchester M13~9PL, United Kingdom}
\affiliation[c]{Department of Physics, University of Cyprus, P.O. Box 20537, 1678 Nicosia, Cyprus}
\abstract{We compute the contributions of the dimension-6 SMEFT operators involving four third-generation quarks to the two-loop renormalisation of the quark fields and masses. We perform the computations in both the Naive Dimensional Regularisation (NDR) and the Breitenlohner-Maison-`t Hooft-
Veltman (BMHV) schemes, and we present results for all relevant field and mass renormalisation constants in the on-shell and \MSbar\ schemes. We carefully discuss all scheme choices which affect the final results. 
This completes the computation of field and mass renormalisation constants relevant for two-loop computations in QCD involving dimension-six SMEFT operators.
}

\begin{document}

\begin{flushright}
BONN-TH-2026-02
\end{flushright}

\maketitle

\section{Introduction}
The Standard Model Effective Field Theory (SMEFT) has been established as a key theoretical framework to parametrise the effects of heavy new physics in the light of the absence of evidence for new light particles. Within the SMEFT, the interactions of the Standard Model (SM) particles are modified by higher-dimensional operators, whose Wilson coefficients are determined by global interpretations of experimental data. 

A systematic effort of improving theoretical predictions in the SMEFT is ongoing, with the aim to maximise the new physics reach of such global SMEFT interpretations. This effort involves the computation of scattering processes at higher orders in the SM couplings, including the automation of next-to-leading order (NLO) QCD corrections \cite{2008.11743, 2507.23549}, progress towards NLO electroweak corrections \cite{1911.11244,2201.09887,Hartmann:2015aia,Hartmann:2015oia,Gauld:2015lmb,Hartmann:2016pil,Dawson:2018dxp,Dawson:2018jlg,Dawson:2018liq,Dawson:2018pyl,Dedes:2018seb,Dedes:2019bew,Cullen:2019nnr,Boughezal:2019xpp,Cullen:2020zof,Corbett:2021cil,Dawson:2021ofa, 2601.15901,2508.14966,2411.08952,2409.11466,2406.03557,
2503.07724,2305.03763,2601.09599} and the first few examples of two-loop computations in the SMEFT and in other EFT frameworks, such as HEFT or LEFT~\cite{1708.00460,1811.12366, Buchalla:2018yce,2202.02333,2204.13045, 2310.18221, 2311.15004, 2410.13304,2409.05728,Braun:2025hvr}. In parallel, efforts continue towards the two-loop renormalisation of the SMEFT at dimension-6 through the computation of the anomalous dimensions at two loops \cite{1910.05831,2005.12917,2308.06315, 2310.19883, 2311.13630,2211.09144, 2203.11224,2401.16904, 2408.03252, 2410.07320,2412.13251, 1907.04923, 2011.02494,Fuentes-Martin:2024agf, 2501.08384, Aebischer:2025zxg, Haisch:2025lvd,Assi:2025fsm,2504.00792,Naterop:2025lzc,2507.08926,2507.10295,2507.19589, 2507.12518,Haisch:2025vqj,Chala:2025crd,2510.08682,2510.14680, 2512.08827}, with the complete set of two-loop anomalous dimensions in Naive Dimensional Regularisation having recently become available~\cite{Born:2026xkr}. 

In the path towards the two-loop renormalisation, the renormalisation of the SM fields and masses is an important ingredient. Steps in this direction have been taken both within LEFT~\cite{2412.13251,2507.08926} and SMEFT \cite{Duhr:2025zqw,2507.10295,Duhr:2025yor}. In previous work \cite{Duhr:2025zqw,Duhr:2025yor}, we computed the two-loop renormalisation of the gluon and quark fields in the presence of the chromomagnetic dipole moment of the top and the triple gluon operators, both in the \MSbar\, and on-shell schemes. We also considered contributions of four-quark operators to the renormalisation of the gluon field and to the running of the strong coupling constant.

Additional components are needed to complete this effort, which also provides a stepping stone towards NNLO computations in the SMEFT. In this work we complete our previous results by computing the renormalisation of the quark fields and masses in both the on-shell and \MSbar\  schemes in the presence of four-quark operators. This extension introduces a significant degree of complexity into the computation. Firstly, the presence of chirality-dependent operators demands particular care related to the treatment of the $\gamma_5$ matrix beyond four dimensions. Secondly, evanescent operators play a crucial role in the computation, due to the fact that spinor identities used to define an operator basis only hold in four dimensions. 

Due to these complications, a series of conventions and schemes have to be chosen, and the final results will depend on these choices. For example, a careful definition of the basis of operators including a set of evanescent operators is required as a starting point of the computation. With respect to the $\gamma_5$ continuation scheme, one can consider both the Naive Dimensional Regularisation (NDR) and the  Breitenlohner-Maison-'t Hooft-Veltman (BMHV) scheme \cite{tHooft:1972tcz, Breitenlohner:1977hr}. The BMHV scheme is the only scheme known to be algebraically consistent to all orders in perturbation theory. For our computation, no ambiguous traces are involved, and so we can also perform the computation in the NDR scheme. For each scheme a set of evanescent operators have to be carefully defined for the calculation. In addition to providing the final results for the renormalisation of the quark fields and masses, we hope that our computation will also serve as a useful guide with regards to all the choices which need to be made and explicitly spelled out before presenting results for such a two-loop computation. 

This work is organised as follows. Our operators are introduced and conventions are set in Section \ref{sec:four-quark-SMEFT}. In Section \ref{sec:two-loopamps} we present one-loop results and discuss both the on-shell and off-shell renormalisation. Results in the NDR-scheme are presented for both the \MSbar\ and on-shell schemes in Section \ref{sec:NDR}, whilst the ones for the BMHV-scheme are discussed in Section \ref{sec: BMHV results}. In Section \ref{sec:comparison}, we compare our findings with existing results from the literature. We conclude in Section \ref{sec: conclusions}. Appendices provide additional information with regards to the renormalisation of the quark masses, evanescent operators and a description of the notebook with the results that we include in the ancillary files.

% !TEX root = main.tex

\section{Four-quark operators of the third generation in the SMEFT}
\label{sec:four-quark-SMEFT}

\subsection{The QCD sector of the SMEFT}

This paper is part of a sequence of works aiming at the two-loop renormalisation of the QCD sector of the dimension-six SMEFT. More specifically, we consider the theory defined by the Lagrangian encompassing the CP-conserving dimension-six operators of the SMEFT that only involve quark and gluon fields,
\beq\label{eq:LQCD6}
\mathcal{L}_{\textrm{QCD},6}= \mathcal{L}_{\text{QCD}}+\frac{c_{G}^0}{\Lambda^2}\,\mathcal{O}_G+\sum_{i,j=1}^{n_u}\frac{c_{uG}^{0,ij}}{\Lambda^2}\mathcal{O}_{uG}^{ij}+\sum_{i,j=1}^{n_d}\frac{c_{dG}^{0,ij}}{\Lambda^2}\mathcal{O}_{dG}^{ij} + \sum_n\frac{c_{4q_n}^{0}}{\Lambda^2}\mathcal{O}_{4q_n}\,,
\eeq 
where $\mathcal{L}_{\text{QCD}}$ is the SM QCD Lagrangian with $n_f=n_u+n_d=6$ different quark flavours ($n_u=3$ and $n_d=3$ are the number of up and down-type quarks, respectively), $\Lambda$ is the SMEFT scale, and $c_{G}^0$, $c_{uG}^{0,ij}$, $c_{dG}^{0,ij}$ and $c_{4q_n}^0$ are the bare Wilson coefficients. This Lagrangian defines the starting point to study the impact of the SMEFT on the renormalisation of quark and gluon fields and masses, as well as of the strong coupling constant.

The operator $\mathcal{O}_G$ is the unique CP-even gauge-invariant operator of dimension six that only involves gluon fields,
\beq\label{eq:triple_gluon_operators}
\mathcal{O}_G = f^{ABC}\,G_{\mu}^{0A\nu}\,G_{\nu}^{0B\rho}\,G_{\rho}^{0C\mu}\,,
\eeq
where $f^{ABC}$ are the SU$(N)$ structure constants and $G_{\mu\nu}^{0A}$ is the (bare) gluon field strength tensor,
\beq
G_{\mu\nu}^{0A} = \partial_{\mu}G_{\nu}^{0A}-\partial_{\nu}G_{\mu}^{0A}-g_s^0\,f^{ABC}\,G_{\mu}^{0B}\,G_{\nu}^{0C}\,,
\eeq
with $G_{\mu}^{0A}$ and $g^0_s$ the bare gluon field and bare strong coupling, respectively.

The chromomagnetic operators in the SMEFT are defined as
\beq\bsp\label{eq:chromo_def_6}
\widetilde{\mathcal{O}}_{uG}^{ij} &\,= {i}\,\overline{Q}^0_iT^A\tau^{\mu\nu}\widetilde{\phi}\,u^0_{Rj}\,G_{\mu\nu}^{0A} + \textrm{h.c.}\,,\qquad 1\le i,j\le n_u\,,\\
\widetilde{\mathcal{O}}_{dG}^{ij} &\,= {i}\,\overline{Q}^0_iT^A\tau^{\mu\nu}{\phi}\,d^0_{Rj}\,G_{\mu\nu}^{0A} + \textrm{h.c.}\,,\qquad 1\le i,j\le n_d\,,
\esp\eeq
where $Q^0_i$ are the bare left-handed quark doublets, $u^0_{Ri}$ and $d^0_{Ri}$ are the bare right-handed quark singlets and  $\phi$ is the SM Higgs doublet, with $\widetilde{\phi} = i\sigma^2\phi^*$. Moreover, $\tau^{\mu\nu} = \frac{1}{2}[\gamma^{\mu},\gamma^{\nu}]$ and $T^A$ are the generators of the fundamental representation SU$(N)$, normalised according to $\Tr(T^AT^B) = \frac{1}{2}\delta^{AB}$. After electroweak symmetry breaking, these operators lead to the following dimension-five operators,
\beq\bsp\label{eq:chromo_def}
\mathcal{O}_{uG}^{ij} &\,= \frac{iv}{\sqrt{2}}\,\overline{u}^0_iT^A\tau^{\mu\nu}u^0_j\,G_{\mu\nu}^{0A}\,,\qquad 1\le i,j\le n_u\,,\\
\mathcal{O}_{dG}^{ij} &\,= \frac{iv}{\sqrt{2}}\,\overline{d}^0_iT^A\tau^{\mu\nu}d^0_j\,G_{\mu\nu}^{0A}\,,\qquad 1\le i,j\le n_d\,,
\esp\eeq
where $v$ is the vacuum expectation value (vev) of the SM Higgs field, and $u_i^0$ and $d_i^0$ are the bare up- and down-type bare quark fields, respectively.
Later on, we will only consider the effect of third-generation quarks, and we introduce the shorthands $\mathcal{O}_{tG} = \mathcal{O}_{uG}^{33}$ and $\mathcal{O}_{bG} = \mathcal{O}_{dG}^{33}$, and the respective Wilson coefficients are $c_{tG}=c_{uG}^{33}$ and $c_{bG}=c_{dG}^{33}$.

A detailed discussion of the operators in eq.~\eqref{eq:LQCD6} can be found in refs.~\cite{Duhr:2025zqw, Duhr:2025yor}. There, we computed the impact of the zero- and two-quark operators on the renormalisation factors of the top-quark and gluon fields, as well as on the renormalisation of the top-quark mass in the \MSbar\ and on-shell renormalisation schemes. We also extracted the renormalisation group (RG) running of the strong coupling constant $\alpha_s$, including the contribution of four-quark operators. The contribution of four-quark operators to the renormalisation constants of the quark masses and fields, however, is still missing to complete this program.
In this paper, we close this gap, and provide the contribution of the four-quark operators to the two-loop renormalisation. 

Considerations from flavour physics restrict the number of allowed four-fermion operators.  From a phenomenological point of view, however, not all four-fermion operators allowed by flavour assumptions are equally relevant. In particular, once we consider current constraints from experimental measurements, four-fermion operators with four heavy quark fields are the least constrained ones.  Thus, they are  particularly interesting from a phenomenological point of view. Motivated by this argument, and also to have a manageable number of operators, we will focus on the following class of four-quark operators involving only third-generation quarks:
 \begin{align}
& 
\left.
\begin{aligned}
  \cO_{QQ}^{(1)} &= \frac{1}{2}(\qbar \gamma^\mu Q)(\qbar \gamma_\mu Q)\,, \\
  \cO_{QQ}^{(8)} &= \frac{1}{2}(\qbar \gamma^\mu T^A Q)(\qbar \gamma_\mu T^A Q)\,,
\end{aligned}
\, \,\right\} (\overline{L}L)(\overline{L}L) \notag \\
& 
\left.
\begin{aligned}
  \cO_{Qt}^{(1)} &= (\qbar \gamma^\mu Q)(\tbar_R \gamma_\mu  t_R)\,, \\
  \cO_{Qt}^{(8)} &= (\qbar \gamma^\mu T^A Q)(\tbar_R \gamma_\mu T^A t_R)\,, \\
  \cO_{Qb}^{(1)} &= (\qbar \gamma^\mu Q)(\bbar_R \gamma_\mu  b_R)\,, \\
  \cO_{Qb}^{(8)} &= (\qbar \gamma^\mu T^A Q)(\bbar_R \gamma_\mu T^A b_R)\,,
\end{aligned}
\, \,\right\} (\overline{L}L)(\overline{R}R) \notag \\
&
\left.
\begin{aligned}
  \cO_{tt} &= (\tbar_R \gamma^\mu t_R)(\tbar_R \gamma_\mu t_R)\,, \\
  \cO_{bb} &= (\bbar_R \gamma^\mu b_R)(\bbar_R \gamma_\mu b_R)\,, \\
  \cO_{tb}^{(1)} &= (\tbar_R \gamma^\mu t_R)(\bbar_R \gamma_\mu b_R)\,, \\
  \cO_{tb}^{(8)} &= (\tbar_R \gamma^\mu T^A t_R)(\bbar_R \gamma_\mu T^A b_R)\,,
\end{aligned}
\, \,\right\} (\overline{R}R)(\overline{R}R) \notag \\
&
\left.
\begin{aligned}
  \cO_{QtQb}^{(1)} &= (\qbar^I  t_R)\epsilon_{IJ}(\qbar^J  b_R)\, + \mathrm{h.c.} \,, \\
  \cO_{QtQb}^{(8)} &= (\qbar^I   T^A t_R)\epsilon_{IJ}(\qbar^J T^A b_R)\, + \mathrm{h.c.} \,,
\end{aligned}
\, \,\right\} (\overline{L}R)(\overline{L}R)
\label{eq: four-quark}
\end{align}
where we have introduced the shorthand notation $\{Q,t_R,b_R\}$\footnote{To simplify the notation in eq.~\eqref{eq: four-quark}, we omit the superscript $^0$ denoting bare quantities.} for the bare SM quark fields of the third generation $\{Q^0_3, u^0_{R3},d^0_{R3}\}$ and $\epsilon_{IJ}$ is the Levi-Civita tensor with two indices.
The basis in eq.~\eqref{eq: four-quark} differs from the Warsaw basis~\cite{1008.4884} by the choice of the $(\overline{L}L)(\overline{L}L)$ operators (operators from the Warsaw basis will be denoted by a superscript `W' in the following). In particular, for the colour-singlet operator we have $\cO_{QQ}^{(1)}=\frac{1}{2}( \cO_{qq}^{(1),{\textrm{W}}})_{3333}$, while the colour octet operator $\cO_{QQ}^{(8)}$ replaces the triplet operator $\cO_{qq}^{(3),{\text{W}}}$. Our basis in eq.~\eqref{eq: four-quark} reproduces the conventions of ref.~\cite{1802.07237} at leading order. We underline that beyond tree-level the definition of the basis becomes subtle, and, for example, defining the octet operator $\cO_{QQ}^{(8)}$ as the octet structure itself, or as a linear combination of a singlet and a triplet operator, will lead to different results for the renormalisation constants and anomalous dimensions, because the two definitions differ by an evanescent operator. We will discuss these subtleties in the next subsection.

Since we are interested in applications to high-energy phenomenology, we consider all quarks from the first two generations massless.
As also discussed in ref.~\cite{Duhr:2025yor}, there is an interplay between the four-quark structures and the assumption of a massless bottom quark, which appears when inserting the four-quark operators into loop diagrams. In particular, the chirality flipping four-fermion operators $\cO^{(R)}_{QtQb}, \,R\in \{1,8\},$ induce a loop-suppressed mass term for the bottom quark if we assume a non-zero mass of the top quark. We observe that this requires special care when performing computations in the five flavour scheme (5FS), which assumes all quarks to be massless, except for the top quark. Therefore, we conclude that higher-order computations in the SMEFT using the 5FS are only consistent if the Wilson coefficients are such that the $2\times2$ mass matrix has a zero eigenvalue. In the following of this paper, we will assume both the top and bottom quarks to be massive. We will comment on the limit of a vanishing bottom mass later on.
Finally, we note that the operators $\cO^{(R)}_{QtQb}$ mix at one-loop with the third-generation up- and down-type chromomagnetic dipole operators. Since these operators contribute to the quark mass counterterm at one-loop \cite{1503.08841, 1607.05330}, this mixing has an impact on the $\cO^{(R)}_{QtQb}$ contribution to the two-loop running of the quark mass. 

\subsection{Basis definition and Fierz-evanescent operators}
When constructing a minimal basis of four-fermion operators, it is customary to use Fierz identities~\cite{Fierz:1939zz} among spinor structures to reduce the number of independent operators. As is well known, these identities only hold in four-dimensional spacetime. For this reason, if we work in dimensional regularisation and wish to use a minimal basis (such as our basis of third-generation four-quark operators in eq.~\eqref{eq: four-quark}, or the more general Warsaw basis~\cite{1008.4884}), it is necessary to take into account that an application of a Fierz identity may leave terms of $\order{\varepsilon}$. These terms can be identified as arising from evanescent operators \cite{Buras:1989xd, Dugan:1990df, hep-ph/9412375, hep-ph/9806471, 2211.09144, 2211.01379, 2208.10513}, which can contribute to the finite parts of one-loop diagrams, and thus to the single UV pole of two-loop diagrams, thereby directly affecting the two-loop RG equations. This implies that the results depend not only on our choice of basis, but also on the specific definition of the operators within that basis. For example the octet $(\bar L L)(\bar L L)$ operator,\footnote{This differs from the octet operator defined in eq.~\eqref{eq: four-quark} by a factor of~2.}
\begin{equation}
  (\mathcal{O}_{qq}^{(8)})_{prst} = (\bar q_p \gamma^\mu T^A q_r)(\bar q_s \gamma_\mu T^A q_t)\,,
\end{equation}
where $q_r$ is the left-handed quark-doublet field from the $r^{\textrm{th}}$ generation, is reduced in the Warsaw basis to a combination of the triplet and singlet operators,
\begin{equation}
  \begin{split}(\mathcal{O}_{qq}^{(3),\textrm{W}})_{prst} &= (\bar q_p \gamma^\mu \tau^I q_r)(\bar q_s \gamma_\mu \tau_I q_t)\,, \\
  (\mathcal{O}_{qq}^{(1),\textrm{W}})_{prst} &= (\bar q_p \gamma^\mu q_r)(\bar q_s \gamma_\mu q_t)\,,
\end{split}
\end{equation}
respectively. Via their relation, we can define
\beq \label{eq:octettotriplet}
(\tilde{\mathcal{O}}_{qq}^{(8),\textrm{W}})_{prst} = \frac{1}{4}\, (\mathcal{O}_{qq}^{(3),\textrm{W}})_{ptsr} + \frac{1}{4}(\mathcal{O}_{qq}^{(1),\textrm{W}})_{ptsr} - \frac{1}{2 N} (\mathcal{O}_{qq}^{(1),\textrm{W}})_{prst}\,.
\eeq
In four dimensions, $\mathcal{O}_{qq}^{(8)}$ and $\tilde{\mathcal{O}}_{qq}^{(8),\textrm{W}}$ are equal, while in $D = 4 - 2\varepsilon$ dimensions they differ by the evanescent operator
\begin{equation}
  (\mathcal{E}_{qq}^{8 \rightarrow 3+1})_{prst} = (\mathcal{O}_{qq}^{(8)})_{prst} - (\tilde{\mathcal{O}}_{qq}^{(8),\textrm{W}})_{prst},
\end{equation}
whose contribution in strictly four dimensions vanishes in the absence of divergences.
%which contributes terms of order $\varepsilon$ when inserted into one-loop diagrams. 

Let us now discuss the definition of our basis in eq.~\eqref{eq: four-quark}. Our choice of operators, together with
the choice of redundant four-quark structures to be eliminated by Fierz
relations, automatically defines a set of evanescent operators. Specifically, given the chosen set of redundant four-quark operators $\{ \mathcal{R}_i \}$, with $i \in \{1, \dots, m\}$, which can be reduced  via Fierz identities to a combination of operators in our basis $\{ \mathcal{O}_j \}$, with $j \in \{1, \dots, n\}$, we define the set of \emph{Fierz-evanescent operators} $\{ \mathcal{E}_i^F \}$ as
\begin{equation}
    \mathcal{E}_i^F = \mathcal{R}_i - \sum_{j=1}^n a_{ij} \, \mathcal{O}_j,
\end{equation}
with $a_{ij}$ the coefficients appearing in the reduction.
The set of relevant evanescent operators defined by our basis in eq.~\eqref{eq: four-quark} (including only operators that do not involve charge-conjugated quark fields, as these are not generated at one loop) is presented in Appendix~\ref{sec: Evanescent operators}. 
%We can write the Lagrangian that collects the Fierz-evanescent operators as
%\begin{equation}\label{eq:L_F_ev}
%    \mathcal{L}_{\mathrm{ev}}^F = \sum_{i}^m \frac{k_{i}^{F,0}}{\Lambda^2} \mathcal{E}_i^F. 
%\end{equation}
%
%Since we are only interested in the quark self-energy at two loops, only a subset of Fierz-evanescent operators contributes to the quark self-energy via one-loop counterterm diagrams, and they all change the chirality of a fermion line. We will comment more on this later in the paper. For now it is suffices to say that only the following Fierz-evanescent operators are relevant for our work:
%\begin{align}\label{eq:evanescentQTQB}
%\mathcal{E}_{QtQb^{(1)}}^F &= \left[(\bar{Q}^I \sigma^{\mu\nu} t)\,\epsilon_{IJ}(\bar{Q}^J \sigma_{\mu\nu} b) + \mathrm{h.c.}\right]
%  + 4\left(1 - \frac{2}{N}\right)\mathcal{O}_{QtQb}^{(1)}
%  - 16\,\mathcal{O}_{QtQb}^{(8)}\,, \\
%\nonumber \mathcal{E}_{QtQb^{(8)}}^F &= \left[(\bar{Q}^I \sigma^{\mu\nu} T^A t)\,\epsilon_{IJ}(\bar{Q}^J \sigma_{\mu\nu} T^A b) + \mathrm{h.c.}\right]
%  - 4\frac{N^2 - 1}{N^2}\mathcal{O}_{QtQb}^{(1)}
%  + 4\left(1 + \frac{2}{N}\right)\mathcal{O}_{QtQb}^{(8)}\,.
%\end{align}

\subsection{The choice of regularisation scheme}
The four-fermion operators in eq.~\eqref{eq: four-quark} distinguish between left- and right-handed components of the fermion fields, even when imposing CP-conservation at the Lagrangian level. A crucial consequence is that when computing amplitudes, we need the chiral projectors 
\begin{equation}
    P_{R/L} = \frac{1 \pm \gamma^5}{2}\,.
\end{equation}
This raises the problem of the treatment of $\gamma^5$ in $D=4-2\varepsilon$ dimensions. This is a well-known issue (see,~e.g., refs.~\cite{hep-th/0005255, 2303.09120}), and a continuation scheme for $\gamma^5$ needs to be adopted. One possible choice for this task is the Naive Dimensional Regularisation (NDR) scheme, which maintains the $\gamma^5$ anti-commutation property with the other gamma matrices in $D$ dimensions. This scheme choice is known to be mathematically inconsistent at higher orders, as it breaks the cyclicity property of the trace. It has, however, been proven to give unambiguous results at one loop, when a consistent reading point for the amplitude is used \cite{Kreimer:1989ke, Korner:1991sx, Kreimer:1993bh, hep-th/0005255, 2012.08506, 2211.09144}. For the computation of the quark self-energy, ambiguous traces involving six ordinary $\gamma$ matrices and one $\gamma^5$ matrix are not encountered, making the NDR-scheme suitable for this specific computation.

Another continuation scheme for the $\gamma^5$ matrix treatment is the Breitenlohner-Maison-'t Hooft-Veltman (BMHV) scheme \cite{tHooft:1972tcz, Breitenlohner:1977hr}. In this scheme, $\gamma^5$ is treated as a purely four-dimensional object, while the $D$-dimensional space-time is split into four- and $(-2\varepsilon)$-dimensional parts. We can then write 
\begin{equation}\label{eq: BMHV def}
    \gamma_\mu = \Bar{\gamma}_\mu + \hat{\gamma}_\mu, \ \ g_{\mu \nu}=\Bar{g}_{\mu \nu}+ \hat{g}_{\mu \nu}\,,
\end{equation}
where we use the symbols $\,\Bar{}\,$ and $\,\hat{}\,$ to indicate the four- and $(-2\varepsilon)$-dimensional components, respectively. In this context, $\gamma^5$ satisfies
\begin{equation}
    \{\Bar{\gamma}_\mu,\gamma_5\}=0 \textrm{~~~and~~~}  [\hat{\gamma}_\mu,\gamma_5]=0\,.
    \label{eq: ga5relations}
\end{equation}
The resulting algebra has been proven to be mathematically consistent at all loop orders \cite{Breitenlohner:1977hr, Speer:1974cz, Breitenlohner:1975hg, Breitenlohner:1976te, Costa:1977pd, Aoyama:1980yw}. In the BMHV-scheme, the vector-type four-quark operators are automatically projected onto the four-dimensional space, because eqs.~\eqref{eq: BMHV def} and \eqref{eq: ga5relations} imply \begin{equation}
\label{eq:vector_proj}
P_{L/R} \, \gamma^\mu P_{R/L} = \bar{\gamma}^\mu \,P_{R/L}\,.
\end{equation}

We will perform the computation in both the NDR and BMHV continuation schemes. It is well known~\cite{Buras:1989xd, hep-ph/9412375, Tracas:1982gp, 2211.09144,2310.13051, Dekens:2019ept} that the adoption of a particular continuation scheme may introduce additional complications due to the appearance of new evanescent structures. Specifically, in the NDR-scheme, some of the identities used to reduce chains of $\gamma$ matrices do not hold in $D$ dimensions, leaving $\order{\varepsilon}$ terms when applied (see, e.g., ref.~\cite{Dekens:2019ept}). Among these, we find that only one relation is relevant to our case, defining the evanescent structure
\begin{equation}\label{eq:NDRevanescent}
    \gamma^\mu \gamma^\nu \gamma^\rho P_{L} \otimes \gamma_\mu \gamma_\nu \gamma_\rho P_{R} - 4 (1+c_{\mathrm{ev}}  \varepsilon)\gamma^\mu P_L \otimes \gamma_\mu P_R.
\end{equation}
The customary choice $c_{\mathrm{ev}} = 1$ ensures that the operator in eq.~\eqref{eq:NDRevanescent} contributes only at $\order{\varepsilon^2}$ when inserted into one-loop diagrams, so that it does not affect the single pole of two-loop diagrams. However, we note that, since we are also interested in the renormalisation factors in the on-shell scheme, the finite parts of the two-loop on-shell diagrams are relevant to our computation, making the structure in eq.~\eqref{eq:NDRevanescent} important nevertheless. For this reason, we perform our computation keeping a generic $c_{\mathrm{ev}}$.

The evanescent operators in the BMHV-scheme do not have the same form as in the NDR-scheme, since the relations for chains of $\gamma$ matrices hold due to the space being split into a four-dimensional and a $(-2\varepsilon)$-dimensional part. 
%We note that Fierz identities still do not hold in the BMHV scheme~[ ]. 
New evanescent structures then arise due to the presence of $\hat{\gamma}$ matrices (see, e.g., ref.~\cite{2310.13051}). The  structures that are generated at one loop and relevant here are discussed in detail in Appendix~\ref{sec: Evanescent operators}, and their impact on the quark self-energy at two loops is discussed in Section~\ref{sec: BMHV results} below. Moreover, in the BMHV-scheme the anomalous dimensions also depend on the precise way the operators are defined. For instance, focusing on the dipole operators from eq.~\eqref{eq:chromo_def}, one may choose either to keep $\tau^{\mu\nu}$ in $D$ dimensions or to consider only its four-dimensional part, $\bar\tau^{\mu\nu}$. These two choices define different regularisation schemes, and they differ by evanescent operators. In this work, we adopt the prescription in which the dimension-six operators are split into a sector which contains purely four-dimensional Dirac structures  and one that is evanescent. Hence, in the BMHV scheme, the operators defined in eqs.~(\ref{eq:triple_gluon_operators})--(\ref{eq: four-quark}) should be understood as having purely four-dimensional Dirac structures.

Another scheme choice concerns the continuation to $D$ dimensions of the renormalisable Lagrangian.  Equation~\eqref{eq:vector_proj} implies that all vector-type fermion bilinears, including the kinetic terms, are projected into four dimensions. It is possible to continue the vector-type bilinears to $D$ dimensions via the inclusion of appropriate evanescent operators.  While this is mandatory for the kinetic terms if we want to employ dimensional regularisation, the continuation of the gluon--fermion interaction is a scheme choice~\cite{2004.14398}.
Keeping this interaction strictly four-dimensional is, in principle, the simplest option, as it generates the smallest number of $\hat{\gamma}$ matrices, and therefore requires handling fewer evanescent operators. However, from a technical perspective, it proved easier to implement the scheme in which the gluon--fermion interaction is continued to $D$ dimensions. Although this leads to a larger number of evanescent structures, these were easier to handle within our workflow. Moreover, this choice of scheme allows us to retain the $SU(3)$ gauge symmetry in $D$ dimensions, and not only in the four-dimensional subspace~\cite{2310.13051}. For these reasons, we perform our computation with a $D$-dimensional gluon--fermion interaction.

Another important consequence of adopting the BMHV scheme is the breaking of chiral symmetry---both global and gauge---when continuing the Lagrangian to $D$ dimensions. In this work, we treat the electroweak sector as static, i.e., $g_1=g_2=0$, with $g_1$ and $g_2$ denoting the $U(1)_Y$ and $SU(2)_L$ gauge couplings, respectively. As a consequence, our computational scheme only breaks a global chiral symmetry. This is less severe than the breaking of a gauge symmetry, for which the inclusion of chiral-symmetry-restoring counterterms is mandatory~\cite{tHooft:1972tcz, Breitenlohner:1977hr, hep-th/9905076, hep-ph/9907426, 2004.14398, 2109.11042, 2205.10381, 2208.09006, 2303.09120, 2406.17013, 2506.12253, 2507.19589}.\footnote{In the unbroken phase of the SMEFT, the gauge chiral-symmetry-restoring counterterms have recently been computed in ref.~\cite{2507.19589}.} In the following, we work in the broken phase of the SM, where chiral symmetry is already broken by the quark mass term. Restoring chiral symmetry with suitable counterterms at this level is possible (as extensively discussed and implemented in refs.~\cite{2310.13051, 2412.13251, 2507.08926}), but it involves introducing non-hermitian mass terms. For this reason, and for simplicity, we refrain from including symmetry-restoring terms for the global chiral symmetry. This defines a different renormalisation scheme from the one used in ref.~\cite{2507.08926}. This needs to be taken into account when comparing our results to existing results in the literature. 

%We collect the evanescent operators generated by the $\gamma^5$-continuation-scheme in the Lagrangian
%\begin{equation}
%    \mathcal{L}_{\mathrm{ev}}^S = \sum_{i}^{N(S)} \frac{k_{i}^{S,0}}{\Lambda^2} \mathcal{E}_i^S, 
%\end{equation}
%with $S \in \{\mathrm{NDR, \mathrm{BMHV}\}}$, and $N(S)$ the number of loop-generated evanescent operators for a given $\gamma^5$-scheme. 
%Putting everything together, we can write the final bare Lagrangian relevant to our work in the form
%\begin{equation}
%    \mathcal{L} =  \mathcal{L}_{\text{QCD},6}  + \mathcal{L}_{\mathrm{II}} + \mathcal{L}_{\mathrm{ev}}^F+ \mathcal{L}_{\mathrm{ev}}^S,
%\end{equation}
%where $\mathcal{L}_{\mathrm{ev}}^F$ was defined in eq.~\eqref{eq:L_F_ev}, and we define $\mathcal{L}_{\mathrm{II}}$ as the Lagrangian collecting all redundant, unphysical (class II) operators that are generated at one-loop and are needed for the renormalisation of off-shell two-loop diagrams. These operators had already been described in ref.~\cite{Duhr:2025zqw}, and we refer to that reference for details.

\subsection{Summary of our choices of regularisation schemes}

The discussions from the preceding sections make it clear that the precise definition of the dimensional regularisation scheme, in particular in the presence of chiral fermions, requires careful definitions and precise settings of conventions. In order to make precise our choices of NDR- and BMHV-schemes, we summarise in this section our definitions and conventions in a concise way (for a similar discussion, see ref.~\cite{Born:2026xkr}).

We start by quoting the precise form of the final bare Lagrangian that we want to renormalise in a $\gamma^5$ continuation scheme $S\in\{\textrm{NDR},\textrm{BMHV}\}$. We can cast it in the form:
\begin{equation}\label{eq:L_final}
    \mathcal{L} =  \mathcal{L}_{\text{QCD},6}  + \mathcal{L}_{\mathrm{II}} + \mathcal{L}_{\mathrm{ev}}^F+ \mathcal{L}_{\mathrm{ev}}^S\,,
\end{equation}
Here $\mathcal{L}_{\text{QCD},6}$ denotes the Lagrangian density from eq.~\eqref{eq:LQCD6}, and $\mathcal{L}_{\mathrm{II}}$ is the Lagrangian collecting all redundant, unphysical (class II) operators that are generated at one-loop and are needed for the renormalisation of off-shell two-loop diagrams. The latter operators had already been described in ref.~\cite{Duhr:2025zqw}, and we refer to that reference for details. The remaining two terms in eq.~\eqref{eq:L_final} contain the evanescent operators. We now describe the two contributions in turn.

The Lagrangian that collects the Fierz-evanescent operators can be cast in the form
\begin{equation}\label{eq:L_F_ev}
    \mathcal{L}_{\mathrm{ev}}^F = \sum_{i=1}^m \frac{k_{i}^{F,0}}{\Lambda^2} \mathcal{E}_i^F. 
\end{equation}
The precise form of these operators depends on the choice of physical operators in $\mathcal{L}_{\textrm{QCD},6}$.
Our complete set of Fierz-evanescent operators defined by eqs.~\eqref{eq: four-quark} and not including charge-conjugated fields is shown in Table~\ref{tab: fierz-evanescent} in Appendix~\ref{sec: Evanescent operators}.
Since we are only interested in the quark self-energy at two loops, only a subset of Fierz-evanescent operators contributes to the quark self-energy via one-loop counterterm diagrams, and they all change the chirality of a fermion line. We will comment more on this later in the paper. For now it suffices to say that only the following Fierz-evanescent operators are relevant for our work:
\begin{align}\label{eq:evanescentQTQB}
\mathcal{E}_{QtQb^{(1)}}^F &= \left[(\bar{Q}^I \sigma^{\mu\nu} t)\,\epsilon_{IJ}(\bar{Q}^J \sigma_{\mu\nu} b) + \mathrm{h.c.}\right]
  + 4\left(1 - \frac{2}{N}\right)\mathcal{O}_{QtQb}^{(1)}
  - 16\,\mathcal{O}_{QtQb}^{(8)}\,, \\
\nonumber \mathcal{E}_{QtQb^{(8)}}^F &= \left[(\bar{Q}^I \sigma^{\mu\nu} T^A t)\,\epsilon_{IJ}(\bar{Q}^J \sigma_{\mu\nu} T^A b) + \mathrm{h.c.}\right]
  - 4\frac{N^2 - 1}{N^2}\mathcal{O}_{QtQb}^{(1)}
  + 4\left(1 + \frac{2}{N}\right)\mathcal{O}_{QtQb}^{(8)}\,,
\end{align}
with $\sigma^{\mu\nu}=i \tau^{\mu\nu}$.

The last term describes evanescent operators related to the specific choice of the $\gamma^5$ continutation scheme $S\in \{\mathrm{NDR, \mathrm{BMHV}\}}$:
\begin{equation}
    \mathcal{L}_{\mathrm{ev}}^S = \sum_{i}^{N(S)} \frac{k_{i}^{S,0}}{\Lambda^2} \mathcal{E}_i^S, 
\end{equation}
with $N(S)$ the number of loop-generated evanescent operators for the given $\gamma^5$-scheme. The precise form of these operators depends on the scheme $S$, and we discuss the NDR- and BMHV-schemes separately in the rest of this section. 

In the NDR-scheme, the operators $\mathcal{E}_i^S$ relevant to us solely arise from the evanescent Dirac structure in eq.~\eqref{eq:NDRevanescent}. The relevant operators are shown in Table~\ref{tab:NDR-evanescent} in Appendix~\ref{sec: Evanescent operators}.

In the BMHV-scheme, we need to consider a larger set of evanescent operators, whose origin can be traced back to the fact that we need to include contributions where ordinary Dirac matrices are replaced by their counterparts $\hat{\gamma}^{\mu}$ in $(-2\varepsilon)$ dimensions. The resulting operators are shown in Tables~\ref{tab:penguin-evanescent} and~\ref{tab:four-quark-evanescent} in Appendix~\ref{sec: Evanescent operators}.

\section{Two-loop self-energy amplitudes}
\label{sec:two-loopamps}
In this section, we discuss our computation and renormalisation procedure for the two-loop self-energy amplitudes. The diagrams have been generated with \texttt{FeynArts} \cite{hep-ph/0012260}, and the Dirac and color algebra, in both the NDR- and BMHV-schemes, have been performed with \texttt{FeynCalc} \cite{2001.04407, 2312.14089, 2512.19858}. The integration-by-parts (IBP) reduction~\cite{chetyrkin:1981qh,Tkachov:1981wb} is performed with \texttt{Kira} \cite{1705.05610, 2008.06494, 2505.20197}. 

\subsection{One-loop results}
Before detailing the two-loop computation, let us recall the one-loop results, which were also partially discussed in ref.~\cite{Duhr:2025yor}. We define the quark mass renormalisation factor in the \MSbar-scheme such that 
\begin{equation}
    m_{\psi}^0 = Z_{m_\psi} \, m_{\psi}\,, \ \ \psi \in \{t, b\}\,,
    \label{eq:Z_m__def}
\end{equation}
where $t$ and $b$ now refer to the physical top and bottom fields, as we work in the broken phase, and $m_{t(b)}^0$ and  $m_{t(b)}$ are the bare and $\overline{\text{MS}}$-renormalised top and bottom masses, respectively. Note that the generic $Z_{m_\psi}$ is a function of both quark masses.\footnote{In ref.~\cite{Duhr:2025yor} we organised the one-loop counterterms as a $2\times2$ mixing matrix between the masses. In this way, one finds non-zero off-diagonal terms stemming from the insertions of $\mathcal{O}_{QtQb}^{(R)}$, with $R\in\{1,8\}$. In particular, this implies that if one starts with a massless bottom quark, a mass-term is generated at higher orders.}
%, effectively breaking the massless assumption.

Given the definition of the mass counterterm in eq.~\eqref{eq:Z_m__def}, we focus on the contribution from the chromomagnetic dipole and four-quark operators, which we indicate with $Z_{m_\psi, qG}$ and $Z_{m_\psi, 4q}$, respectively. Due to the one-loop mixing between $\mathcal{O}_{QtQb}^{(R)}$ and the dipole operators, the one-loop contribution of the latter to the quark mass is relevant for the two-loop running contribution from $\mathcal{O}_{QtQb}^{(R)}$. We can write the expansion,
\begin{equation}\begin{split}\label{eq:ZmExpansion}
    Z_{m_\psi,4q}&\, = \frac{1}{16 \pi^2 \Lambda^2}\sum_{\ell\geq 1} \left(\frac{\alpha_s}{4 \pi} \right)^{\ell-1}\delta Z_{m_\psi,4q}^{(\ell)}\,, \\
    Z_{m_\psi,qG} &\,= \frac{1}{16 \pi^2 \Lambda^2}\sum_{\ell\geq 1} \left(\frac{\alpha_s}{4 \pi} \right)^{\ell-1}\delta Z_{m_\psi,qG}^{(\ell)}\,,\end{split}
\end{equation}
with the one-loop term taking the form
\begin{equation}
\begin{split}
    \delta Z_{m_t,4q}^{(1)} & =
  -\frac{4 m_t^2}{\varepsilon}\qty(c_{Qt}^{(1)}+\CF\, c_{Qt}^{(8)}) + \frac{m_b^3}{m_t \, \varepsilon}\left((2 \CA+1)\, c_{QtQb}^{(1)}+\CF\, c_{QtQb}^{(8)}\right)\,, \\ \delta Z_{m_b,4q}^{(1)} & =-\frac{4 m_b^2}{\varepsilon}\qty(c_{Qb}^{(1)}+\CF\, c_{Qb}^{(8)}) +\frac{m_t^3}{m_b \, \varepsilon}\left((2 \CA+1)\, c_{QtQb}^{(1)}+\CF \,c_{QtQb}^{(8)}\right)\,,  \\  \delta Z_{m_t,qG}^{(1)} & = \frac{12}{\sqrt{2} \,\varepsilon} \, c_{tG}\, g_s  \,m_t\,v\,, \qquad  \delta Z_{m_b,qG}^{(1)}  = \frac{12}{\sqrt{2} \,\varepsilon} \, c_{bG}\, g_s \,m_b\,v\,.
\end{split}
\label{eq: mass mixing matrix}
\end{equation}
The Wilson coefficients $c_{QtQb}^{(R)}$ with $R\in\{1,8\}$ are real as we are interested in the CP-conserving Lagrangian, and we introduced the usual values of the quadratic Casimir operators of the fundamental and adjoint representations of SU$(N)$,
\beq
\CF = \frac{N^2-1}{2N}\textrm{~~~and~~~} \CA = N\,.
\eeq
Due to their chiral structure, the four-quark operators distinguish between left- and right-handed components of the quark fields, and we define the left- and right-handed field renormalisation factors as 
\begin{equation}\label{eq:ZpsiMS}
    \psi^0_L= \sqrt{(Z_\psi^L)_{\overline{\textrm{MS}}}}\, \psi_L\,, \qquad \psi^0_R= \sqrt{(Z_\psi^R)_{\overline{\textrm{MS}}}}\, \psi_R\,,
\end{equation}
with $\psi \in \{t,b\}$, and the superscript $^0$ indicates the bare quantities.
Similarly to eq.~\eqref{eq:ZmExpansion}, focusing on the four-quark operators, we can write the expansion
\begin{equation}\label{eq:Zpsi_expansion}
     (Z_{\psi,4q}^\chi)_{\overline{\textrm{MS}}} = \frac{1}{16 \pi^2 \Lambda^2}\sum_{\ell\geq 1} \left(\frac{\alpha_s}{4 \pi} \right)^{\ell-1}\Big(\delta Z_{\psi,4q}^{\chi,(\ell)}\Big)_{\overline{\textrm{MS}}}\,,
\end{equation}
with $\chi\in\{L,R\}$. We note that the field renormalisation factors do not receive any corrections at one loop from the CP-even four-fermion operators, implying $\delta Z_{\psi,4q}^{\chi,(1)}=0$. 

In the on-shell scheme we define the quark mass renormalisation factors as
\begin{equation}
    m_\psi^0 = Z_{M_\psi} M_\psi,\,  \ \ \psi \in \{t, b\}\,,
\end{equation}
with $M_t$ and $M_b$ the on-shell renormalised masses of the top and bottom quarks, respectively. %As the on-shell renormalisation factor includes finite terms, we can already observe differences between the computations in the NDR- and BMHV-schemes. 
We can expand the on-shell mass renormalisation factor for a given regularisation scheme $S\in\{\textrm{NDR},\textrm{BMHV}\}$ as
\begin{equation}\begin{split}\label{eq:ZM_expansion}
    Z_{M_\psi, 4q}^S &\,= \frac{1}{16 \pi^2 \Lambda^2}\sum_{\ell\geq 1} \left(\frac{\alpha_s}{4 \pi} \right)^{\ell-1}\delta Z_{M_\psi,4q}^{S,(\ell)}\,, \\ Z_{M_\psi, qG}^S &\,= \frac{1}{16 \pi^2 \Lambda^2}\sum_{\ell\geq 1} \left(\frac{\alpha_s}{4 \pi} \right)^{\ell-1}\delta Z_{M_\psi,qG}^{S,(\ell)}\,.
    \end{split}
\end{equation}
The one-loop results are
\begin{align}
   \nonumber   \delta Z_{M_t,4q}^{(1),S} &= C(\varepsilon, M_t) \,4 M_t^2 \delta_S^{4q}(\varepsilon)\frac{c_{Qt}^{(1),0}+\CF c_{Qt}^{(8),0}}{ \varepsilon(\varepsilon-1)} - C(\varepsilon, M_b) \, \frac{M_b^3}{M_t} \frac{(2 \CA+1)c_{QtQb}^{(1),0}+\CF c_{QtQb}^{(8),0}}{  \varepsilon (\varepsilon-1)}\,, \\ 
   \nonumber\delta Z_{M_b,4q}^{(1),S}& = C(\varepsilon, M_b) \,4 M_b^2 \delta_S^{4q}(\varepsilon)\frac{c_{Qb}^{(1),0}+\CF c_{Qb}^{(8),0}}{\varepsilon(\varepsilon-1)} - C(\varepsilon, M_t) \, \frac{M_t^3}{M_b} \frac{(2 \CA+1)c_{QtQb}^{(1),0}+\CF c_{QtQb}^{(8),0}}{  \varepsilon (\varepsilon-1)}\,, \\
     \delta Z_{M_t,qG}^{(1),S}&= C(\varepsilon, M_t) \frac{12 g_s\, c_{tG}^0 M_t \, v}{\sqrt{2}\, \varepsilon(1-\varepsilon)} \delta_S^{qG}(\varepsilon)\,, \\ \delta Z_{M_b,qG}^{(1),S}&= C(\varepsilon, M_b) \frac{12 g_s\, c_{bG}^0 M_b \, v}{\sqrt{2}\, \varepsilon(1-\varepsilon)} \delta_S^{qG}(\varepsilon)\,,\nonumber
\end{align}
where $C(\varepsilon, M)$ is a normalisation factor defined as,
\begin{equation}
    C(\varepsilon,M)=\Gamma(1+\varepsilon) (4 \pi)^\varepsilon M^{-2\varepsilon}\,,
    \label{eq: Ceps constant}
\end{equation}
and $\delta_S(\varepsilon)$ takes the form
\begin{equation}
 \delta_S^{4q}(\varepsilon) = 
\begin{cases}
1, & S = \text{BMHV},\\[6pt]
1 - \dfrac{\varepsilon}{2}, & S = \text{NDR}\,,
\end{cases}
\qquad
 \delta_S^{qG}(\varepsilon) = 
\begin{cases}
\dfrac{3-\varepsilon}{3-2\varepsilon}, & S = \text{BMHV},\\[6pt]
\dfrac{3- 2\varepsilon}{3}, & S = \text{NDR}\,.
\end{cases}
\end{equation}
We also define the on-shell renormalisation factors for the left- and right-handed quark fields as
\begin{equation}\label{eq:ZpsiOS}
    \psi^0_L= \sqrt{\left(Z_\psi^L\right)_{\mathrm{OS}}}\, \psi_L^\mathrm{OS}\,, \qquad \psi^0_R= \sqrt{\left(Z_\psi^R\right)_{\mathrm{OS}}}\, \psi_R^\mathrm{OS}\,.
\end{equation}
Similarly to eq.~\eqref{eq:Zpsi_expansion}, we can write the expansion
\begin{equation}
    \left(Z_{\psi,4q}^\chi\right)_{\mathrm{OS}} = \frac{1}{16 \pi^2}\sum_{\ell\geq 1} \left(\frac{\alpha_s}{4 \pi} \right)^{\ell-1}\left(\delta Z_{\psi,4q}^{\chi,(\ell)}\right)_{\mathrm{OS}},
\end{equation}
where at one-loop $\left(\delta Z_{\psi,4q}^{\chi,(1)}\right)_{\mathrm{OS}}=0$.

\subsection{Off-shell reduction and renormalisation}
\begin{figure}[htbp]
\centering
\includegraphics[width=0.75\linewidth]{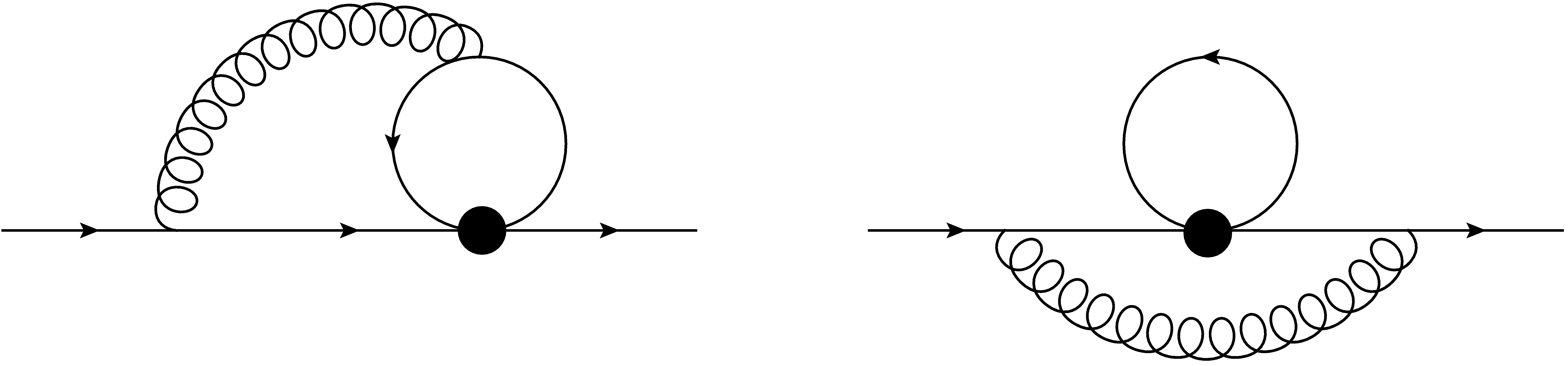}
\caption{Two-loop example diagrams contributing to the quark self-energy with an insertion of a four-quark operator, represented by the black dot.}
\label{fig:tt_4F_direct}
\end{figure}

Let us focus on the \MSbar-scheme, which requires the computation of the self-energies to be performed in an off-shell environment. The loop integrals contributing to the top-quark self-energy (see, e.g, Figure~\ref{fig:tt_4F_direct}) at two loops can be mapped onto the two integral families shown in Table~\ref{tab: top_families}. The families relevant to the computation of the bottom self-energy are obtained by replacing $m_t$ with $m_b$ and vice-versa.

\renewcommand{\arraystretch}{1.2}
\begin{table}[ht]
    \centering
        \begin{tabular}{c|c}
          $T_A$ &  $T_B$ \\ \hline
          $k_1^2$ & $k_1^2$ \\
          $k_2^2-m_t^2$ & $k_2^2-m_b^2$ \\
          $(k_1-p)^2-m_t^2$ & $(k_1-p)^2-m_t^2$\\
          $(k_2-p)^2$ & $(k_2-p)^2$\\
          $(k_1-k_2)^2-m_t^2$ & $(k_1-k_2)^2-m_b^2$
    \end{tabular}
    \caption{Definition for the set of integral families required for the calculation of the top-quark self-energy at two loops with a massive bottom quark. }
    \label{tab: top_families}
\end{table} 
\noindent The master integrals (MIs) relevant to our computation are
\begin{equation}
I^{T_A}_{01100}\,,I^{T_A}_{01101},\,I^{T_A}_{01102},\,I^{T_A}_{11100},\,
I^{T_B}_{01001}\,,I^{T_B}_{01100},\,I^{T_B}_{11100},\,I^{T_B}_{01101},\, I^{T_B}_{01102},\,I^{T_B}_{01201}.
\end{equation}
As we are interested in the renormalisation factors in the \MSbar-scheme, we only retain the poles of these integrals, which are known in the literature~\cite{hep-ph/0501132}, and avoid the complication of elliptic integrals, especially due to the presence of sunrise integrals with different masses, cf.,~e.g.,~refs.~\cite{Caffo:1998du,Laporta:2004rb,Adams:2014vja,Adams:2015gva,Adams:2017ejb,Broedel:2017siw,1203.6543}.

The renormalisation procedure is analogous to the one discussed in refs.~\cite{Duhr:2025zqw,Duhr:2025yor}, where we computed one-loop counterterm diagrams with the insertion of both physical and redundant operators. This requires the one-loop off-shell renormalisation of both the quark-quark-gluon vertex (via penguin diagrams) and the four-quark vertex. 
A novel feature of our computation compared to our previous works~\cite{Duhr:2025yor,Duhr:2025zqw} is the presence of evanescent operators, that are generated at one loop and need to be inserted in one-loop counterterm diagrams. Due to their dependence on the $\gamma^5$-continuation scheme, the  evanescent operators will be discussed in dedicated sections.

As we have already noted that at one loop, the four-quark operators in eq.~\eqref{eq: four-quark} do not contribute to the renormalisation factors of the quark fields. At two loops, we find the first contribution of these operators to $Z_{\psi}^\chi$.
Redundant, unphysical, class-II operators are also generated at two loops. Their definition is
\begin{equation}\label{eq:redundant-2-point}
   \mathcal{O}^{L/R}_{\partial^3\psi}= \bar \psi (\Box+m_\psi^2)i\slashed{\partial}\,P_{L/R} \psi\,, \qquad \mathcal{O}_{\partial^2\psi}= \bar \psi (i\slashed{\partial}-m_\psi)^2\,\psi\,.
\end{equation}

Let us now comment on our specific choice of \MSbar\ renormalisation scheme. Specifically, since we work in the presence of evanescent operators, their insertions into loop diagrams can lead to finite mixing into physical operators. To cancel this effect, we adopt a renormalisation scheme in which finite counterterms are added to the usual \MSbar \ renormalisation factors. These counterterms depend on the coefficients of evanescent operators and are fixed by the finite mixing into physical operators. This is a customary choice~\cite{Buras:1989xd, Dugan:1990df, hep-ph/9412375, 2310.13051}, which is known to change the two-loop RG equations. We cast the counterterm for the mass as
\begin{equation}\label{eq:Zm_modMS}
    Z_{m_\psi}^{S} = (Z_{m_\psi}^{S})_{\overline{\mathrm{MS}}}  + \sum_j \frac{k_j^S}{\Lambda^2} \delta m_{\psi, \mathrm{ev}, j}^{S} + \sum_i \frac{k_i^F}{\Lambda^2} (\delta m_{\psi, \mathrm{ev},i}^{F})^S\,,
\end{equation}
where $\delta m_{\psi, \mathrm{ev},j}^{S}$ and $(\delta m_{\psi, \mathrm{ev},i}^{F})^S$ are the finite counterterms relative to insertions of evanescent operators of Fierz- and $\gamma^5$-continuation-type, respectively. We also observe that the insertion of evanescent operators of Fierz-type depends on the $\gamma^5$-scheme, which motivates the additional $S$ superscript in $(\delta m_{\psi, \mathrm{ev}}^{F})^S$.

\subsection{On-shell reduction and renormalisation}
In order to extract the renormalisation factors in the on-shell scheme, we impose the on-shell renormalisation conditions on the fermionic two-point function. In general, it can be written as
\begin{equation}
    \Gamma^\psi(p^2, m_t^0, m_b^0) = i\,\slashed{p} \left(P_L \Sigma_L^\psi(p^2, m_t^0, m_b^0) + P_R \Sigma_R^\psi(p^2, m_t^0, m_b^0)\right)+ i\,m_\psi^0 \Sigma_S^\psi(p^2, m_t^0, m_b^0)\,,
\end{equation}
with $\psi \in \{t,b\}$, and where we can extract the form factors $\Sigma_{L,R}^\psi$ and $\Sigma_S^\psi$ via suitable projectors.
We have already introduced the on-shell renormalisation factors for the quark masses $Z_{M_\psi}$ and for the right- and left-handed quark fields $\left(Z_\psi^{L,R}\right)_{\mathrm{OS}}$. 
Let us define the combination of form factors
\begin{equation}
    \tilde\Sigma^\psi(p^2, m_t^0, m_b^0)=\Sigma_L^\psi(p^2, m_t^0, m_b^0) +\Sigma_R^\psi(p^2,  m_t^0, m_b^0)+2Z_{M_\psi}\Sigma_S^\psi(p^2,  m_t^0, m_b^0)\,.
\end{equation}
The renormalisation factors can then be extracted from $\tilde{\Sigma}^{\psi}$ via~\cite{0709.1075}:
\begin{align}\label{eq:On-shell ren cond}
& Z_{M_\psi} = 1+\frac{1}{2}\eval{\tilde\Sigma^\psi(p^2, Z_{M_t} M_t, Z_{M_b} M_b)}_{p^2=M_\psi^2}, \\ \nonumber & \left(Z_\psi^{L}\right)_{\mathrm{OS}}^{-1}=\eval{\Sigma_L^\psi(p^2,  Z_{M_t} M_t, Z_{M_b} M_b)}_{p^2=M_\psi^2}+ p^2 \eval{\frac{\partial}{\partial p^2}\tilde\Sigma^\psi(p^2, Z_{M_t} M_t, Z_{M_b} M_b)}_{p^2=M_\psi^2}, \\
\nonumber&
\left(Z_\psi^{R}\right)_{\mathrm{OS}}^{-1}=\eval{\Sigma_R^\psi(p^2,  Z_{M_t} M_t, Z_{M_b} M_b)}_{p^2=M_\psi^2}+ p^2 \eval{\frac{\partial}{\partial p^2}\tilde\Sigma^\psi(p^2, Z_{M_t} M_t, Z_{M_b} M_b)}_{p^2=M_\psi^2}.
\nonumber
\end{align}
Equation~\eqref{eq:On-shell ren cond} can be solved perturbatively in the loop expansion. We note that at the two-loop order, this gives rise to additional terms involving derivatives with respect to the mass that need to be included for a proper treatment. The on-shell field renormalisation constants include the insertion of the one-loop counterterm of the class-II penguin operator into the one-loop self-energy.

The on-shell limit is performed at the level of the integrand, and also the IBP reduction is performed on-shell, organising the amplitude into the same integral families as shown in Table~\ref{tab: top_families}. The resulting MIs (which are also on-shell) read
\begin{equation}
    I^{T_A}_{01100}, \, I^{T_A}_{01101}, \, I^{T_A}_{10011}, \, I^{T_B}_{01001}, \, I^{T_B}_{01100}, \, I^{T_B}_{01101},\,I^{T_B}_{01102}\,. 
\end{equation}
Their expressions are known in the literature~\cite{hep-ph/9304303, hep-ph/9712209}. We note that we have a reduced number of MIs with respect to the off-shell reduction.

\section{Results in the NDR-scheme}
\label{sec:NDR}
In this section, we present the results for the renormalisation factors computed in the NDR-scheme. We performed the computation in a general gauge, retaining the explicit dependence on the gauge parameter $\xi$. The $\xi$-independence of the mass renormalisation factor serves as a check of our computation.

\subsection{Evanescent operators}
We now discuss the insertion of evanescent operators in the NDR-scheme. As previously explained in Section~\ref{sec:four-quark-SMEFT}, we can distinguish two different classes of evanescent operators, namely those arising from the Fierz identities (see Table~\ref{tab: fierz-evanescent}) and those appearing due to the $\gamma^5$-continuation scheme (see Tables~\ref{tab:NDR-evanescent}, \ref{tab:penguin-evanescent} and \ref{tab:four-quark-evanescent}).

We start by discussing the insertion of Fierz-evanescent operators into one-loop diagrams in the NDR-scheme. We only have four-quark evanescent operators that are generated at one loop, and they can only contribute to the mass counterterm via tadpole integrals. The bare contribution of the Fierz-evanescent operator to the top-quark self-energy is given by
\begin{equation}\label{eq:FierzEvanescent_Contribution}
\begin{split}
   & \expval{\mathcal{E}_{QtQb}^F}_{1L}^{\mathrm{NDR}} = \frac{C(\varepsilon,m_b)}{8 \pi^2 \Lambda^2} \frac{m_b^3}{m_t}\frac{7-2\varepsilon}{\varepsilon-1}\left(k_{QtQb^{(1)}}^{0,F}+C_F k_{QtQb^{(8)}}^{0,F}\right)\,.
\end{split}
\end{equation}
Note that this expression is finite in the limit $\varepsilon \rightarrow 0$. In the convention of eq.~\eqref{eq:Zm_modMS}, we can extract from eq.~\eqref{eq:FierzEvanescent_Contribution} the finite counterterms $(\delta m_{\psi, \rm{ev}, i}^F)^{\rm{NDR}}$, with $i\in\{QtQb^{(1)}, QtQb^{(8)}\}$.

We now discuss the implications of the Fierz-evanescent operators on the \MSbar-scheme. The Wilson coefficients $k_{QtQb^{(R)}}^{0,F}$ of the Fierz-evanescent operators are loop-suppressed, and the result is proportional to the Wilson coefficients of our physical basis, due to the mixing of our physical operators into the evanescent structures. This leads to the two-loop contribution
\begin{equation}\label{eq:FierzContributionMSbar}
\eval{\expval{\mathcal{E}_{QtQb}^F}_{1L}^{\mathrm{NDR}}}_{\overline{\mathrm{MS}}} = \frac{7 C_F \, g_s^2}{512 \pi^4 \Lambda^2 \varepsilon} \frac{m_b^3}{m_t}\left(4 c_{QtQb}^{(1)}+(4 C_F- C_A) c_{QtQb}^{(8)}\right),
\end{equation}
which directly contributes a simple pole to the quark mass counterterm.

Next, we focus on the operators obtained from the structures in eq.~\eqref{eq:NDRevanescent}. After dressing the Dirac matrices in eq.~\eqref{eq:NDRevanescent} with quark fields, we obtain colour-singlet and colour-octet operators, which contribute to the top-quark self-energy as
\begin{equation}\label{eq:NDRev_contribution}
    \expval{\mathcal{E}_{LR}^{(3)}}^{\mathrm{NDR}}_{1L} = \frac{C(\varepsilon,m_t)}{2 \pi^2} m_t^2 \frac{(2-\varepsilon)(c_{\mathrm{ev}}-1+\varepsilon)}{1-\varepsilon}\left(k_{LR^{(3)},t}^{(1),0}+C_F k_{LR^{(3)},t}^{(8),0}\right),
\end{equation}
where the additional superscript on the Wilson coefficients of the evanescent operators indicates the colour representation of the operators generated by the evanescent Dirac structure in eq.~\eqref{eq:NDRevanescent}. Similar to eq.~\eqref{eq:FierzContributionMSbar}, we find 
\begin{equation}\label{eq:NDRev_contribution_MS}
\eval{\expval{\mathcal{E}_{LR}^{(3)}}^{\mathrm{NDR}}_{1L}}_{\overline{\mathrm{MS}}} = \frac{C_F \, g_s^2}{64 \pi^4 \Lambda^2 \varepsilon} m_t^2 \,(c_{\mathrm{ev}}-1)\left(4 c_{Qt}^{(1)} + (4 C_F-C_A)c_{Qt}^{(8)}\right),
\end{equation}
which vanishes for $c_{\mathrm{ev}}=1$. Similarly, to eq.~\eqref{eq:FierzEvanescent_Contribution}, eq.~\eqref{eq:NDRev_contribution} allows us to extract the finite counterterms to the mass of the quark in the NDR scheme, $\delta m_{\psi, \rm{ev}, j}^{\rm{NDR}}$ (see eq.~\eqref{eq:Zm_modMS}).
We note that these are the only evanescent contributions to our computation in the NDR-scheme, because the tadpole diagram only receives contributions from chirality-flipping four-fermion operators such as $\mathcal{O}_{Qt}^{(R)}$ and $\mathcal{O}_{QtQb}^{(R)}$. Indeed, at one loop, additional evanescent operators are generated in the NDR-scheme, but they are of the form $(\bar L L)(\bar L L)$ or $(\bar R R)(\bar R R)$, and they do therefore not contribute to our computation. We remark here that the choice of the renormalisation scheme of eq.~\eqref{eq:Zm_modMS}, which includes finite counterterms proportional to the renormalised evanescent coefficients, is such that it cancels the mixing of such operators in the physical sector, as in eqs.~\eqref{eq:FierzEvanescent_Contribution} and \eqref{eq:NDRev_contribution}. As a result, the evanescent operators only contribute via their counterterms which are proportional to the physical coefficients, as in eqs.~\eqref{eq:FierzContributionMSbar} and \eqref{eq:NDRev_contribution_MS}.

\subsection{Running of the quark masses} 
We now focus on the \MSbar\ counterterms. Let us start with the wave-function renormalisation. We can write the $\ell$-loop four-quark contribution to the wave-function renormalisation (see eq.~\eqref{eq:Zpsi_expansion}) as
\begin{equation}\label{eq:wave-function-exp-MS}
     \delta Z_{\psi, 4q}^{\chi,(\ell)}= \sum_{k=-\ell}^{-1} C_F\, \varepsilon^k \delta Z_{\psi, 4q}^{\chi, (\ell, k)},
\end{equation}
with $\chi \in \{L, R\}$. Since the four-quark operators do not contribute at one loop, we have $\delta Z_{\psi,4q}^{L,(1,-1)} = \delta Z_{\psi,4q}^{R,(1,-1)} = 0$. The first non-vanishing contribution appears at two loops. 
In the following, we only present the results for the top-quark renormalisation factors. Owing to the symmetry of our setup, the corresponding results for the bottom-quark field are immediately obtained by applying the following set of replacements:
\begin{equation}\label{eq:mtmb-replacement}
    m_t \leftrightarrow m_b, \qquad c_{Qt}^{(R)} \leftrightarrow c_{Qb}^{(R)},\qquad c_{tt}\leftrightarrow c_{bb}\,,
\end{equation}
which we will denote with shorthand notation $t \leftrightarrow b$.
The expressions for the field renormalisation factors then read

\begin{align} 
    \delta Z_{t,4q}^{L, (2,-2)} &= \frac{1}{3}\, c_{Qb}^{(8)} m_t^2
+ \frac{2}{3}\, c_{QQ}^{(1)} m_t^2
- \frac{ C_A - 2 C_F-2}{3}\, c_{QQ}^{(8)} m_t^2 \notag\\&
+ \frac{1}{3}\, c_{Qt}^{(8)} m_t^2
- \frac{3}{2}\, m_b m_t\, c_{QtQb}^{(1)}
+ \frac{3}{4}\,(C_A - 2 C_F)\, m_b m_t\, c_{QtQb}^{(8)}\,,  \notag\\
    \left(\delta Z_{t,4q}^{L, (2,-1)}\right)_{\mathrm{NDR}} &= \frac{2}{3}\, c_{Qb}^{(8)} m_t^2
+ 2\, c_{QQ}^{(1)} m_t^2
+ 3\, c_{Qt}^{(1)} m_t^2
+ \frac{1}{2}\, c_{tb}^{(8)} m_t^2
+ 2\, c_{tt}^{(1)} m_t^2 \notag \\ &
- \frac{ 3 C_A - 6 C_F-4}{3}\, c_{QQ}^{(8)} m_t^2
- \frac{ 9 C_A - 18 C_F-10}{6}\, c_{Qt}^{(8)} m_t^2 \notag\\&
+ \frac{19}{4}\, m_b m_t\, c_{QtQb}^{(1)}
- \frac{19}{8}\,(C_A - 2 C_F)\, m_b m_t\, c_{QtQb}^{(8)}\,,  \label{eq:wave-function-MS-NDR}\\
    \delta Z_{t,4q}^{R, (2,-2)} &=  \frac{2}{3}\, c_{Qt}^{(8)} m_t^2
+ \frac{1}{3}\, c_{tb}^{(8)} m_t^2
+ \frac{4}{3}\, c_{tt}^{(1)} m_t^2 \notag\\&
- \frac{3}{2}\, m_b m_t\, c_{QtQb}^{(1)}
+ \frac{3}{4}\,(C_A - 2 C_F)\, m_b m_t\, c_{QtQb}^{(8)}\,, \notag\\
    \left(\delta Z_{t,4q}^{R, (2,-1)}\right)_{\mathrm{NDR}} &= \frac{1}{2}\, c_{Qb}^{(8)} m_t^2
+ \, c_{QQ}^{(1)} m_t^2
+ 3\, c_{Qt}^{(1)} m_t^2
+ \frac{2}{3}\, c_{tb}^{(8)} m_t^2
+ 4\, c_{tt}^{(1)} m_t^2 \notag\\ &
- \frac{ C_A - 2 C_F-2}{2}\, c_{QQ}^{(8)} m_t^2
- \frac{ 9 C_A - 18 C_F-11}{6}\, c_{Qt}^{(8)} m_t^2 \notag\\&
+ \frac{19}{4}\, m_b m_t\, c_{QtQb}^{(1)}
- \frac{19}{8}\,(C_A - 2 C_F)\, m_b m_t\, c_{QtQb}^{(8)}\,,
\nonumber
\end{align}
and accordingly for the bottom quark. We did not label the double pole with $(\cdot)_{\mathrm{NDR}}$, as it is independent of the chosen $\gamma^5$-scheme. We also remark that the expressions in eq.~\eqref{eq:wave-function-MS-NDR} take a gauge-invariant form, being independent of the gauge parameter $\xi$.

A distinctive feature of the \MSbar-scheme compared to the on-shell scheme is that the quark masses satisfy a renormalisation group equation (RGE) in the \MSbar-scheme.
The RGE for the quark masses in the NDR-scheme reads
\begin{equation}
    \frac{\dd m_\psi}{\dd\! \log \mu} = \left(\gamma_{m_\psi}\right)_{\mathrm{NDR}} \, m_\psi\,,
\end{equation}
where the anomalous dimension $\left(\gamma_{m_\psi}\right)_{\mathrm{NDR}}$ is obtained from the mass counterterm matrix in eq.~\eqref{eq:ZmExpansion} together with the mixing term into evanescent operators which contribute to the renormalisation of the mass \cite{2310.13051}. The full expression of the mass counterterms can be found in Appendix~\ref{sec:Appendix_renormalisation_factors}. We note that the coefficient of the double pole is fixed by the consistency and finiteness conditions of the RGEs, and in this case we are required to obtain a gauge-invariant result.
We can expand the contribution from the four-quark operators to the mass anomalous dimension as
\begin{equation}\label{eq:mass_RGE_exp}
    \gamma_{m_\psi,4q}=\frac{C_F}{16 \pi^2 \Lambda^2}\sum_{\ell\geq1} \left(\frac{\alpha_s}{4\pi}\right)^{(\ell-1)}  \gamma_{m_\psi,4q}^{(\ell)}\,,
\end{equation}
with the two-loop coefficient given by 
\begin{align}\label{eq:mass_RGE_ndr}
\left(\gamma_{m_t,4q}^{(2)} \right)_{\mathrm{NDR}}  =& \, \frac{64}{3} c_{QQ}^{(1)} m_t^2
+ \frac{128}{3} c_{tt}^{(1)} m_t^2
+ \frac{4}{3} c_{Qb}^{(8)} (9 m_b^2 + m_t^2)
+ \frac{4}{3} c_{tb}^{(8)} (9 m_b^2 + m_t^2) \notag\\ &
+ \frac{4}{3} c_{QQ}^{(8)} \left[9 m_b^2 + (11+16\CF-8\CA)m_t^2\right]
+ 8 c_{Qt}^{(1)} m_t^2 (4 c_{\mathrm{ev}} - 5) \notag\\ &
+ 4 c_{Qt}^{(8)} \left[
3 m_b^2 + m_t^2\left(7 - 10\CF-7\CA - 2 c_{\mathrm{ev}}\left(\CA-4\CF\right)\right)
\right] \notag\\  &
+ 2 \frac{m_b}{m_t} c_{QtQb}^{(1)}\left(m_b^2(8\CA + 22)+7 m_t^2\right)  \notag\\ &
- \frac{m_b}{m_t}c_{QtQb}^{(8)}\left[11 m_b^2(\CA-4\CF)+7 m_t^2(\CA-2\CF)\right], \notag\\
\left(\gamma_{m_b,4q}^{(2)} \right)_{\mathrm{NDR}} =  &\eval{\left(\gamma_{m_t,4q}^{(2)} \right)_{\mathrm{NDR}} }_{t \leftrightarrow b}.
\end{align}
\subsection{The on-shell scheme} \label{sec:on-shell-NDR}
We now turn our attention to the on-shell scheme. Using the relations in eq.~\eqref{eq:On-shell ren cond}, we determine the renormalisation factors. The two-loop contribution can be expanded as
\begin{equation}
\begin{split}
    \left(\delta Z_{\psi,4q}^{\chi, (2)}\right)_{\mathrm{OS}}& =C(\varepsilon,M_\psi)^2 \sum_{k=-2}^0 C_F\,\varepsilon^k \, \left(\delta Z_{\psi,4q}^{\chi, (2, k)}\right)_{\mathrm{OS}}, \\ \delta Z_{M_\psi,4q}^{(2)}&=C(\varepsilon,M_\psi)^2 \sum_{k=-2}^0 C_F \,\varepsilon^k \, \delta Z_{M_\psi,4q}^{(2,k)}\,,
\end{split}
\label{eq:on-shell-2L-expansion}
\end{equation}
where the normalisation factor $C(\varepsilon, M_\psi)$ was defined in eq.~\eqref{eq: Ceps constant}. The explicit expressions are rather lengthy and contain (di)logarithms of the ratio of the top- and bottom-quark masses. For this reason, we present here only the result for the top-quark on-shell mass counterterm proportional to the coefficients $c_{Qt}^{(1)}$ and $c_{Qt}^{(8)}$, as it is representative of the generic structure. The full set of results is provided as a computer-readable ancillary file with the arXiv submission. We note that the on-shell field and mass renormalisation factors are independent of the gauge parameter $\xi$, which serves as a check of our computation. 

 Focusing only on the terms dependent on the bare $c_{Qt}^{(1),0}$ and $c_{Qt}^{(8),0}$, the top-quark mass counterterm reads
\begin{align}\label{eq:ZmOS-cQt8}
    \delta Z_{M_t,4q}^{(2,-2)} = & \,12 M_t^2c_{Qt}^{(1),0}+ 2 M_t^2 c_{Qt}^{{(8),0}}\,(6 \CF+1) + \dots, \notag \\
    \left(\delta Z_{M_t,4q}^{(2,-1)}\right)_{\mathrm{NDR}} = & \, 40 M_t^2c_{Qt}^{(1),0}+ M_t^2 c_{Qt}^{{(8),0}}\,(40 \CF- 9 \CA + 9) + 3 M_b^2 c_{Qt}^{{(8),0}} + \dots, \notag\\
    \left(\delta Z_{M_t,4q}^{(2,0)}\right)_{\mathrm{NDR}}= &\, 85 M_t^2c_{Qt}^{(1),0}+\frac{M_t^2 c_{Qt}^{{(8),0}}}{18}\left(1530 \CF-351 \CA -28 \pi^2+423\right)\notag\\ & -\frac{M_b^2 c_{Qt}^{(8),0}}{72}(20 \pi^2-441)+c_{Qt}^{{(8)},0} F_{c_{Qt}^{(8)}}(M_t,M_b) + \dots, 
\end{align}
where the dots indicate terms proportional to Wilson coefficients of other operators. The function $F_{c_{Qt}^{(8)}}(M_t,M_b)$ takes the form
\begin{align} \label{eq:F_cQt8}
    F_{c_{Qt}^{(8)}}(M_t,M_b)=& - \frac{4 M_t^2}{3}\left[2\log^2 \left(\frac{M_t}{M_b}\right)+\mathrm{Li}_2 \left(1-\frac{M_t^2}{M_b^2}\right)\right] \notag\\ & + M_b^2\left[\frac{8}{3}\log \left(\frac{M_t}{M_b}\right)+\frac{161}{24}+\frac{5\pi^2}{18}\right] - \frac{16 M_b^3}{3 M_t}\left[\Xi\left(\frac{M_b}{M_t}\right)+\frac{\pi^2}{3}\right] \notag\\ & + \frac{4 M_b^4}{3 M_t^2}\left[2\log \left(\frac{M_t}{M_b}\right)-1\right] + \frac{2 M_b^6}{9 M_t^4}\left[\pi^2-6\,\mathrm{Li}_2 \left(1-\frac{M_t^2}{M_b^2}\right)\right]\,,
\end{align}
where $\Xi(x)=\mathrm{Li}_2(1-x)+\mathrm{Li}_2(-x)+\log x\,\log(1+x)$ is real for $x>0$, and we introduced the dilogarithm function,
\beq
\textrm{Li}_2(z) = \sum_{k=1}^\infty\frac{z^k}{k^2}\,.
\eeq
The function $F_{c_{Qt}^{(8)}}$ originates from the expressions of the two-loop on-shell integrals with two different masses. An important feature is that it is independent of the $\gamma^5$-scheme, since it appears for the first time at two-loop order and does not involve any rational terms associated with one-loop subgraphs. Similar scheme-independent functions also arise from other four-quark operators. Specifically, the terms proportional to $c_{Qb}^{(8)}$, $c_{tb}^{(8)}$, $c_{QQ}^{(8)}$, $c_{QtQb}^{(1)}$, and $c_{QtQb}^{(8)}$ exhibit the same behaviour as those proportional to $c_{Qt}^{(8)}$. In contrast, the terms proportional to $c_{QQ}^{(1)}$ and $c_{tt}$ behave like $c_{Qt}^{(1)}$, with a purely rational contribution that depends only on $M_t^2$.

Let us conclude by commenting on the limit of a vanishing bottom-quark mass, which is a setup often employed when performing computations for a hadron collider (the so-called 5-flavour scheme). We have already mentioned that, even if we start from a massless bottom quark, a mass term will generically be generated radiatively from four-quark operators. In such a scenario it is inconsistent to work with a massless bottom quark, unless the effect of the operators that generate the mass cancel. This is in particular the case if the contributions from $\mathcal{O}_{QtQb}^{(R)}$ vanish. In that case, for instance, the finite term of eq.~\eqref{eq:ZmOS-cQt8} becomes
\begin{equation}
    \eval{\left(\delta Z_{M_t,4q}^{(2,0)}\right)_{\mathrm{NDR}}}_{M_b\rightarrow0} = 85 M_t^2c_{Qt}^{(1),0}+\frac{c_{Qt}^{(8),0} M_t^2}{6}\left(510\, \CF-117\,\CA-8 \pi^2+141 \right)+\dots,
\end{equation}
where in particular we used
\begin{equation}
    \eval{F_{c_{Qt}^{(8)}}(M_t,M_b)}_{M_b\rightarrow0} = \frac{2}{9}M_t^2\pi^2.
\end{equation}

\subsection{The relation between the $\overline{\textrm{MS}}$ and on-shell masses}
\label{sec:MSOSRelation_NDR}
With both the \MSbar\ and on-shell results at hand, we can compute the relation between the renormalised masses in the two schemes. We can write the relation between the on-shell and \MSbar\ renormalised masses as
\begin{equation}\label{eq:MSOSmassRelation}
    M_{\psi} = \frac{Z_{m_\psi}}{Z_{M_\psi}} m_\psi\,,
\end{equation}
with $m_\psi=m_\psi(\mu)$ the renormalised \MSbar\ mass. The renormalisation factors in eq.~\eqref{eq:MSOSmassRelation} depend on the renormalised masses of the top and bottom quarks and on the bare and renormalised couplings, including the Wilson coefficients. To write a consistent relation, we shift all the parameters to their \MSbar\ renormalised counterparts, for example we write the top-quark mass on-shell renormalisation factor as
\begin{equation}
    Z_{M_t} = Z_{M_t}(M_t, M_b, c_i^0, g_s^0)=Z_{M_t}\left(\frac{Z_{m_t}}{Z_{M_t}} m_t,\frac{Z_{m_b}}{Z_{M_b}} m_b, Z_{ij}c_j, Z_{g_s} g_s\right).
\end{equation}
Equation~\eqref{eq:MSOSmassRelation} is then expanded in the strong coupling constant. The ratio is finite, which serves as an additional check for our computation. Focusing on the contribution from the four-quark operators, we can write 
\begin{equation}
    \left(\frac{Z_{m_\psi}}{Z_{M_\psi}}\right)_{4q} =  \frac{1}{16 \pi^2 \Lambda^2} \sum_{\ell\geq1} \left(\frac{\alpha_s}{4\pi}\right)^{(\ell-1)}\xi_{\psi,4q}^{(\ell)}\,.
\end{equation}
For the top-quark, the one-loop term takes a simple form and reads
\begin{equation}\label{eq:ZMOS_1L}
    \left(\xi_{t,4q}^{(1)}\right)_{\mathrm{NDR}} =  2 m_t^2\left(c_{Qt}^{(1)}+\CF c_{Qt}^{(8)}\right)- \frac{m_b^3}{m_t} \left[2\log \left(\frac{m_t}{m_b}\right)+1\right]\left((2\CA+1)c_{QtQb}^{(1)}+\CF c_{QtQb}^{(8)}\right).
\end{equation}
The two-loop term is more involved, and following what we did for the on-shell renormalised mass, we only show the result proportional to $c_{Qt}^{(1)}$ and $c_{Qt}^{(8)}$, and we present the rest only in electronic form. It reads
\begin{align}
    \left(\xi_{t,4q}^{(2)}\right)_{\mathrm{NDR}} = & \,\CF\,m_t^2 c_{Qt}^{(1)}\left(29-8 c_{\mathrm{ev}}\right)+\frac{\CF m_t^2\, c_{Qt}^{(8)}}{18}\left[36 c_{\mathrm{ev}}(\CA-4\CF)+387\CA +522\CF \right. \notag \\ & \left. +28 \pi^2-387\right]+ \frac{\CF m_b^2\, c_{Qt}^{(8)}}{72}(20 \pi^2-441)-\CF  \, c_{Qt}^{(8)} F_{c_{Qt}^{(8)}}(m_t,m_b) +\dots
\end{align}

\section{Results in the BMHV-scheme} \label{sec: BMHV results}
In this section, we present our results in the BMHV-scheme. This scheme choice is not unique, and additional prescriptions are required when continuing the Lagrangian to $D$ dimensions. These have been discussed in Section~\ref{sec:four-quark-SMEFT}.

\subsection{Evanescent operators}
Let us now discuss the insertion of evanescent operators in one-loop diagrams. 
Similarly to the NDR-scheme, we also have in the BMHV-scheme the Fierz-evanescent operators of eq.~\eqref{eq:evanescentQTQB}. 
However, their contribution vanishes,
\begin{equation}
    \expval{\mathcal{E}_{QtQb}^F}_{1L}^{\mathrm{BMHV}}
    = 0 \, .
\end{equation}
A similar vanishing was also observed in ref.~\cite{2310.13051}.
The computation in the BMHV-scheme requires a larger set of evanescent operators. 
While the relevant evanescent operators in the NDR-scheme were limited to four-quark structures, 
in the BMHV-scheme we also encounter quark--gluon evanescent operators generated by one-loop penguin diagrams. 
We have already discussed these in the context of the two-loop gluon self-energy in ref.~\cite{Duhr:2025yor}, where they only mix within the evanescent sector and are therefore negligible at two-loop order. 
For the quark self-energy, however, such operators mix into the physical sector, and their contributions must be taken into account.

Focusing on the insertion of the fermion–gluon evanescent operators in the one-loop diagrams, we find that they generate a finite term proportional to all the tensor structures appearing in the quark self-energy. The resulting expression is rather lengthy, and provided in the \texttt{Mathematica} notebook included in the ancillary files, so we only quote the final contribution to the quark-mass counterterm. For the top-quark case, it reads
\begin{equation}\label{eq:HVevanescent2F}
\begin{split}
    (\delta m_t)_{\mathcal{E}_{\psi^2 G}} = & \,  C(\varepsilon, m_t) \frac{g_s m_t \CF}{4 \pi^2 \Lambda^2 (\varepsilon-1)(2\varepsilon-3)}\left[\frac{1}{2}(\varepsilon-5)(k_{tG}^{L,0}+k_{tG}^{R,0})+m_t(3-2\varepsilon)\,k_{tGD}^{LR,0} \right. \\ & \left. +(1+2 \varepsilon)\,k_{tG}^{LR,0}\right],
\end{split}
\end{equation}
where the Wilson coefficients of the evanescent operators are defined by the operators listed in Table~\ref{tab:penguin-evanescent} of Appendix~\ref{sec: Evanescent operators}. Note that the result for the bottom quark is simply obtained by replacing $m_t$ with $m_b$ and the superscript `$tG$' and `$tGD$' with `$bG$' and `$bGD$', which is equivalent to replacing the top-quark field with the bottom-quark field in the definition of the evanescent operators.

Similarly, we consider the insertion of the four-fermion evanescent operators in the one-loop diagrams, and we denote the final contribution to the top-quark mass counterterm by $(\delta m_t)_{\mathcal{E}_{\psi^4}}$. It reads
\begin{align}
    (\delta m_t)_{\mathcal{E}_{\psi^4}} = & -C(\varepsilon, m_t)\frac{m_t^2}{4 \pi^2 \Lambda^2 (\varepsilon-1)}\left[ k_{S, RR, tt}^{(1),0}+\CF k_{S, RR,tt}^{(8),0} +12 \left(k_{T, RR,tt}^{(1),0}+ \CF k_{T, RR,tt}^{(8),0}\right) \right. \notag \\ & \left. +4(\varepsilon+1)\left(k_{V,RL,tt}^{(1),0} + \CF k_{V,RL,tt}^{(8),0}\right)\right] \notag \\ & +C(\varepsilon, m_b)\frac{m_b^3}{4 \pi^2 \Lambda^2 \,m_t (\varepsilon-1)}\left[ (1+\varepsilon)\left(k_{S, RL,tb}^{(1),0}+\CF k_{S, RL,tb}^{(8),0}\right)+\CA k_{S, RL,tb}^{(1),0} \right. \notag \\ & \left. +2 \left(k_{V, RL,tb}^{(1),0}+ \CF k_{V, RL, tb}^{(8),0}\right) \right]\,. \label{eq:HVevanescent4F}
\end{align}
The result for the bottom quark is again obtained by replacing $m_t$ with $m_b$ and the `$tt$' subscript in the first two lines of eq.~\eqref{eq:HVevanescent4F} with `$bb$'. The corresponding Wilson coefficients of evanescent operators are defined by the operators listed in Table~\ref{tab:four-quark-evanescent} of Appendix~\ref{sec: Evanescent operators}.

We observe that, due to the one-loop mixing of physical operators into the evanescent sector, eqs.~\eqref{eq:HVevanescent2F} and \eqref{eq:HVevanescent4F} give rise to a single pole at two loops proportional to the Wilson coefficients of physical operators. 

\subsection{Running of the quark masses}
We now discuss the results in the \MSbar-scheme (modified by the inclusion of finite counterterms due to the insertions of evanescent operators in loop diagrams). Focusing on the wave-function renormalisation, the expansion takes the same form as in eq.~\eqref{eq:wave-function-exp-MS}. Again we will only show the results for the top quark, as the ones for the bottom quark can be obtained by the usual replacement rules. The double pole is independent of the chosen $\gamma^5$-scheme and identical to its expression in the NDR-scheme in eq.~\eqref{eq:wave-function-MS-NDR}. Therefore, we only show the single pole in the BMHV-scheme, which reads:
\begin{align}
    \left(\delta Z_{t,4q}^{L, (2,-1)}\right)_{\mathrm{BMHV}} &= \frac{5}{6}\, c_{Qb}^{(8)} m_t^2
+ \frac{1}{3}\, c_{QQ}^{(1)} m_t^2
+ \frac{4}{9}\, c_{Qt}^{(1)} m_t^2
+ \frac{5}{9}\, c_{tb}^{(8)} m_t^2
- \frac{40}{9}\, c_{tt}^{(1)} m_t^2 \notag\\ &
- \frac{ C_A - 2 C_F-10 }{6}\, c_{QQ}^{(8)} m_t^2
- \frac{ 4 C_A - 8 C_F-35 }{18}\, c_{Qt}^{(8)} m_t^2 \notag\\ &
+ \frac{87 m_b - 8 m_t}{36}\, m_t\, c_{QtQb}^{(1)}
- \frac{(C_A - 2 C_F)(87 m_b - 8 m_t)}{72}\, m_t\, c_{QtQb}^{(8)}\,, \\
\left(\delta Z_{t,4q}^{R, (2,-1)}\right)_{\mathrm{BMHV}} & = \frac{5}{9}\, c_{Qb}^{(8)} m_t^2
- \frac{20}{9}\, c_{QQ}^{(1)} m_t^2
+ \frac{4}{9}\, c_{Qt}^{(1)} m_t^2
+ \frac{5}{6}\, c_{tb}^{(8)} m_t^2
+ \frac{2}{3}\, c_{tt}^{(1)} m_t^2 \notag\\ &
+ \frac{10(C_A - 2 C_F+1)}{9}\, c_{QQ}^{(8)} m_t^2
- \frac{2(C_A - 2 C_F-10)}{9}\, c_{Qt}^{(8)} m_t^2 \notag\\ &
+ \frac{87 m_b - 8 m_t}{36}\, m_t\, c_{QtQb}^{(1)}
- \frac{(C_A - 2 C_F)(87 m_b - 8 m_t)}{72}\, m_t\, c_{QtQb}^{(8)}\,.
\label{eq:ZtBMHV}
\end{align}

We show the two-loop expressions for the counterterms for the quark mass in Appendix \ref{sec:Appendix_renormalisation_factors}. For the running of the quark mass, we have
\begin{equation}
    \frac{\dd m_\psi}{\dd \log \mu} = \left(\gamma_{m_\psi}\right)_{\mathrm{BMHV}} \,m_\psi\,.
\end{equation}
Expanding the anomalous dimension as in eq.~\eqref{eq:mass_RGE_exp}, we get
\begin{align}
\left(\gamma_{m_t,4q}^{(2)} \right)_{\mathrm{BMHV}}  =& \, 40\, c_{QQ}^{(1)} m_t^2
+ 80\,c_{tt}^{(1)} m_t^2
+12 c_{Qb}^{(8)} m_b^2
+ 12 c_{tb}^{(8)} m_b^2 \notag \\ &
+ 4 c_{QQ}^{(8)} \left(3 m_b^2 + (3+10\CF-5\CA)m_t^2\right)
-80  c_{Qt}^{(1)} m_t^2  \notag\\ &
+ 4 c_{Qt}^{(8)} \left(
3 m_b^2 + 2 m_t^2\left(3-10 \CF
\right)\right) \notag\\  &
+ 2 \frac{m_b}{m_t} c_{QtQb}^{(1)}\left(4 m_b^2(2\CA + 7)+ m_t^2\right) \notag \\ &
- \frac{m_b}{m_t}c_{QtQb}^{(8)}\left(2 m_b^2(9\CA-28\CF+2)+ m_t^2(\CA-2\CF)\right),\notag \\
\left(\gamma_{m_b,4q}^{(2)} \right)_{\mathrm{BMHV}} =  &\eval{\left(\gamma_{m_t,4q}^{(2)} \right)_{\mathrm{BMHV}} }_{t \leftrightarrow b}\,. \label{eq:mass_RGE_BMHV}
\end{align}

\subsection{The on-shell scheme}
We now turn our attention to the results in the on-shell scheme with our BMHV continuation scheme for $\gamma^5$. In particular, let us comment on the on-shell limit employed in the renormalisation conditions of eq.~\eqref{eq:On-shell ren cond}. In the BMHV-scheme, $p^2=\bar p^2+\hat{p}^2$ as the $D$ dimensional space-time is split into a $4$- and $(-2\varepsilon)-$dimensional part. As the on-shell limit projects onto the physical space of the particle, we consider the on-shell condition to hold in four dimensions, where we set the $(-2\varepsilon)-$dimensional part to be vanishing, i.e., $\bar p^2 = M_\psi^2$ and $\hat{p}^2 = 0$. 

Similarly to the NDR-scheme discussed in Section~\ref{sec:on-shell-NDR}, we expand the renormalisation factors as in eq.~\eqref{eq:on-shell-2L-expansion}. The two-loop result for the top-quark mass renormalisation factor in the on-shell scheme reads 
\begin{align}\label{eq:ZmOS-cQt8-HV}
    \left(\delta Z_{M_t,4q}^{(2,-1)}\right)_{\mathrm{BMHV}} = & \,\, 20 M_t^2 c_{Qt}^{{(1),0}}+  2 M_t^2 c_{Qt}^{{(8),0}}\,(10 \CF+ 2 \CA + 5) + 3 M_b^2 c_{Qt}^{{(8),0}} + \dots, \notag\\
    \left(\delta Z_{M_t,4q}^{(2,0)}\right)_{\mathrm{BMHV}}= &\, \frac{160}{9} M_t^2 c_{Qt}^{{(1),0}}+\frac{2 M_t^2 c_{Qt}^{{(8),0}}}{9}\left(80 \CF+95 \CA -7 \pi^2+123\right)\notag\\&-\frac{M_b^2 c_{Qt}^{(8),0}}{72}(20 \pi^2-513)+c_{Qt}^{(8),0} F_{c_{Qt}^{(8)}}\left(M_t, M_b\right)+ \dots, & 
\end{align}
where we again did not show the double pole result as it is identical to the NDR-scheme, the dots indicate terms proportional to other Wilson coefficients, and the function $F_{c_{Qt}^{(8)}}\left(M_t, M_b\right)$  is defined in eq.~\eqref{eq:F_cQt8}. All the other results are shown in electronic format in the ancillary files. 

We can also extract the relations between the \MSbar\ and on-shell renormalised quark masses. The one-loop result is identical to the NDR-scheme (see eq.~\eqref{eq:MSOSmassRelation}), while the two-loop result, proportional to the Wilson coefficients $c_{Qt}^{(R)}$, reads
\begin{align}
     \left(\xi_{t,4q}^{(2)}\right)_{\mathrm{BMHV}} = & \,\CF\,m_t^2 \frac{1036}{9}c_{Qt}^{(1)}+\frac{2\CF m_t^2\, c_{Qt}^{(8)}}{9}\left(518\CF-25\CA+7\pi^2-105\right)\notag\\ &+ \frac{\CF m_b^2\, c_{Qt}^{(8)}}{72}(20 \pi^2-513)-\CF \, c_{Qt}^{(8)} F_{c_{Qt}^{(8)}}(m_t,m_b) +\dots\,.
\end{align}

\section{Comparison with existing results}
\label{sec:comparison}
In ref.~\cite{2507.08926}, the two-loop running of the quark masses in the presence of four-fermion operators appears as part of the general two-loop renormalisation of the Low-Energy Effective Theory (LEFT). Because of the structural differences between SMEFT and LEFT, the four-fermion operators are defined differently. Nevertheless, since we work in the broken phase, we can map our operators in eq.~\eqref{eq: four-quark} to those used in ref.~\cite{2507.08926}, which follow the conventions of ref.~\cite{2310.13051}. Specifically, we have
\begin{align}\label{eq:SMEFTtoLEFT_map}
 & \mathcal{O}_{Qt}^{(1)} = \mathcal{O}_{uu}^{V1,LR}+ \mathcal{O}_{du}^{V1,LR}, \qquad \mathcal{O}_{Qt}^{(8)} = \mathcal{O}_{uu}^{V8,LR}+ \mathcal{O}_{du}^{V8,LR}, \qquad \mathcal{O}_{tt} = \mathcal{O}_{uu}^{V1,RR}, \notag \\
 & \mathcal{O}_{Qb}^{(1)} = \mathcal{O}_{dd}^{V1,LR}+ \mathcal{O}_{ud}^{V1,LR}, \qquad \mathcal{O}_{Qb}^{(8)} = \mathcal{O}_{dd}^{V8,LR}+ \mathcal{O}_{ud}^{V8,LR}, \qquad \mathcal{O}_{bb} = \mathcal{O}_{dd}^{V1,RR}, \notag \\
 & 2\mathcal{O}_{QQ}^{(1)} = \mathcal{O}_{uu}^{V1,LL}+ \mathcal{O}_{dd}^{V1,LL}+2\mathcal{O}_{ud}^{V1,LL},   \qquad  \mathcal{O}_{tb}^{(1)} = \mathcal{O}_{ud}^{V1,RR}, \qquad \mathcal{O}_{tb}^{(8)} = \mathcal{O}_{ud}^{V8,RR}, \notag \\ & \mathcal{O}_{QtQb}^{(1)} = \mathcal{O}_{ud}^{S1,RR} - \mathcal{O}_{uddu}^{S1,RR} + \mathrm{h.c.}, \qquad \mathcal{O}_{QtQb}^{(8)} = \mathcal{O}_{ud}^{S8,RR} - \mathcal{O}_{uddu}^{S8,RR} + \mathrm{h.c.},
\end{align}
where the LEFT operators in the right-hand side of eq.~\eqref{eq:SMEFTtoLEFT_map} are evaluated for the third generation for the up- and down-quark. 
The mapping of $\mathcal{O}_{QQ}^{(8)}$ to the operators to the LEFT basis is not straightforward, as it requires additional Fierz-evanescent operators, because the $(\bar L L)(\bar L L)$ octet operator is reduced via Fierz identities. Therefore, we refrain from doing so and we do not perform this comparison here.
%We note that the operator $\mathcal{O}_{QQ}^{(8)}$ cannot be mapped to the LEFT operators, since the octet $(\bar L L)(\bar L L)$ structure is removed via Fierz identities from the LEFT basis. 
The results in ref.~\cite{2507.08926} are computed in a BMHV-scheme similar to the one used in this work to compute the anomalous dimension in eq.~\eqref{eq:mass_RGE_BMHV}. However, a key difference in the scheme adopted in ref.~\cite{2507.08926} is the inclusion of chiral-symmetry-restoring finite counterterms.
We used the chiral-symmetry-restoring finite counterterms computed in ref.~\cite{2310.13051}, together with the mapping in eq.~\eqref{eq:SMEFTtoLEFT_map}, to obtain the anomalous dimension of the quark mass in the scheme employed in ref.~\cite{2507.08926}, which we denote as `$\chi \mathrm{BMHV}$'. For example, for the top quark, it reads:

\begin{align}
\left(\gamma_{m_t,4q}^{(2)} \right)_{\chi \mathrm{BMHV}}  =& \, \frac{176}{9} c_{QQ}^{(1)} m_t^2
+ \frac{352}{9} c_{tt}^{(1)} m_t^2
+ \frac{4}{9} c_{Qb}^{(8)} (27 m_b^2 - 5 m_t^2)
+ \frac{4}{9} c_{tb}^{(8)} (27 m_b^2 - 5 m_t^2) \notag \\ &
-64 c_{Qt}^{(1)} m_t^2  
+ \frac{4}{3} c_{Qt}^{(8)} \left(
9 m_b^2 + m_t^2\left(13-48 \CF\right)
\right) \notag\\  &
+ 2 \frac{m_b}{m_t} c_{QtQb}^{(1)}\left(4 m_b^2(5\CA + 4)+3 m_t^2\right)  \notag \\ &
- \frac{m_b}{m_t}c_{QtQb}^{(8)}\left(4 m_b^2(2\CA-8\CF+1)+3 m_t^2(\CA-2\CF)\right)\,,
\end{align}
which agrees with ref.~\cite{2507.08926}.%except for the term in red.

Results in the NDR-scheme are available from ref.~\cite{2507.10295} for operators that involve four top-quark fields, namely $\mathcal{O}_{Qt}^{(1)}, \mathcal{O}_{Qt}^{(8)}, \mathcal{O}_{tt}, \tilde{\mathcal{O}}_{QQ}^{(1)},  \tilde{\mathcal{O}}_{QQ}^{(3)}$. With respect to the definitions given in eq.~\eqref{eq: four-quark}, we have $\tilde{\mathcal{O}}_{QQ}^{(1)} = 2\mathcal{O}_{QQ}^{(1)}$ and the triplet $\tilde{\mathcal{O}}_{QQ}^{(3)}= (\mathcal{O}_{qq}^{(3),\mathrm{W}})_{3333}$ is preferred to the octet operator $\mathcal{O}_{QQ}^{(8)}$. Given this starting basis, our results for the running of the top-quark mass in the NDR-scheme in eq.~\eqref{eq:mass_RGE_ndr} agree with those of ref.~\cite{2507.10295} for the overlapping operator set when taking $c_{\mathrm{ev}}=1$. In the context of the NDR-scheme, ref.~\cite{Born:2026xkr} recently presented the full two-loop anomalous dimension matrix in the unbroken phase of the theory. We verified that, after accounting for the relation between the quark masses and the Yukawa couplings in the SMEFT (see, e.g., ref.~\cite{1312.2014}), our results agree with those of ref.~\cite{Born:2026xkr}.
Reference~\cite{2507.10295} also proposes a method to convert the results from the NDR- to the BMHV-scheme based on the mapping of ref.~\cite{DiNoi:2025uan} and without the need  of a proper two-loop computation in the BMHV-scheme. As discussed in earlier sections, there are various conventions required to define a precise variant of BMHV-scheme, in particular how evanescent operators are defined. The numerous differences between the BMHV scheme employed in this work and that of ref.~\cite{DiNoi:2025uan} make a detailed comparison technically involved and therefore beyond the scope of this paper. Since our results agree with the BMHV computation from ref.~\cite{2507.08926} (after accounting for the differences in the precise definition of the scheme discussed above), we are confident that our results are correct. 

%We were not able to identify the precise version of the BMHV-scheme which should be obtained via the conversion from the NDR-scheme, rendering a direct comparison with our results difficult. In particular, we were not able find agreement with the results in the BMHV-type scheme from ref.~\cite{2507.10295}
\section{Conclusions}
\label{sec: conclusions}
In this work we computed the contribution of the dimension-6 four-quark operators to the renormalisation of the quark fields and masses at two loops. Our computation was  performed in both the on-shell and \MSbar\ schemes. Due to the presence of chirality-dependent operators, the treatment of $\gamma_5$ beyond four dimension adds an additional layer of complexity to the computation. We employ both the algebraically consistent BMHV-scheme and the NDR-scheme. Special care is needed to define the operator basis including the set of evanescent operators entering the calculation in both schemes. Scheme choices play a central role in the computation and can affect the final results, therefore we have discussed these in detail. While results in the \MSbar-scheme had already been available, our results in the on-shell scheme are new and presented here for the first time.

Our results complete the computation of the two-loop renormalisation of fields within the QCD sector of the dimension-6 SMEFT. Our calculation complements our previous work on the contributions of the chromomagnetic and triple gluon operators on the renormalisation of quark and gluon fields and the running of the strong coupling constant, as well as the contributions of four-quark operators to the gluon wavefunction. This set of results is a vital ingredient for the computation and subsequent  phenomenological studies of processes involving quarks and gluons in the SMEFT at two loops. 

\section*{Acknowledgements}
We are grateful to Peter Stoffer for help with the comparison to the results of ref.~\cite{2507.08926}, and also for a careful reading of and comments on the manuscript. We also thank Ramona Gr\"ober for comments on the comparison with ref.~\cite{2507.10295}.
GV and EV are supported by the European Research Council (ERC) under the European
Union’s Horizon 2020 research and innovation programme (Grant agreement No. 949451) and by a Royal Society University Research Fellowship through grant URF/R1/201553. The work of CD is supported by the DFG project 499573813 “EFTools”.

\appendix
\section{Renormalisation factors for the quark mass}
\label{sec:Appendix_renormalisation_factors}
In this section we show the full expressions for the \MSbar \, renormalisation factors for the masses of the quarks both in NDR- and BMHV-schemes, $(Z_{m_\psi}^S)_{\text{\MSbar}}$ in eq~\eqref{eq:Zm_modMS}. Focusing on the two-loop contribution from four-quark operators (see eq.~\eqref{eq:ZmExpansion}), we expand the mass counterterm as
\begin{equation}
    \delta Z_{m_t,4q}^{(2)} = \sum_{k=-2}^{-1}C_F\,\varepsilon^k \delta Z_{m_t,4q}^{(2,k)}\,.
\end{equation}
The results for the top-quark read:
\begin{align}
   \delta Z_{m_t,4q}^{(2,-2)} = & -\frac{4}{3}\,c_{QQ}^{(1)}\,m_t^{2}
-\frac{8}{3}\,c_{tt}\,m_t^{2}
-\frac{2}{3}\,c_{Qb}^{(8)}\,m_t^{2}
-\frac{2}{3}\,c_{tb}^{(8)}\,m_t^{2} \notag \\ &
+\frac{2}{3}\,c_{QQ}^{(8)}\!\left(C_A-2(1+C_F)\right)m_t^{2}
+36\,c_{Qt}^{(1)}\,m_t^{2}\notag \\ &
+2\,c_{Qt}^{(8)}\!\left(-1+18C_F\right)m_t^{2}\notag 
-3\,c_{QtQb}^{(1)}\,\frac{m_b}{m_t}\Bigl((5+6C_A)m_b^{2}+m_t^{2}\Bigr)
\notag \\ &+\frac{3}{2}\,c_{QtQb}^{(8)}\,\frac{m_b}{m_t}\Bigl(C_A(2m_b^{2}+m_t^{2})-2C_F(5m_b^{2}+m_t^{2})\Bigr)\,, \notag \\ 
 \left(\delta Z_{m_t,4q}^{(2,-1)}\right)_{\mathrm{NDR}} = &\frac{16}{3}\,c_{QQ}^{(1)}\,m_t^{2}
+\frac{32}{3}\,c_{tt}\,m_t^{2}
+\frac{1}{3}\,c_{Qb}^{(8)}\!\left(9m_b^{2}+m_t^{2}\right)
+\frac{1}{3}\,c_{tb}^{(8)}\!\left(9m_b^{2}+m_t^{2}\right)  \notag \\ &
+c_{QQ}^{(8)}\!\left(
3m_b^{2}+\frac{(11-8C_A+16C_F)m_t^{2}}{3}
\right)
+2\,c_{Qt}^{(1)}\!\left(-9+8c_{\mathrm{ev}}\right)m_t^{2}  \notag \\ &
+c_{Qt}^{(8)}\!\left(
3m_b^{2}
+\bigl(7-(5+4c_{\mathrm{ev}})C_A+2(-9+8c_{\mathrm{ev}})C_F\bigr)m_t^{2}
\right)  \notag \\ &
+c_{QtQb}^{(1)}\!\left(
\frac{2(9+2C_A)m_b^{3}}{m_t}
+\frac{7}{2}m_b m_t
\right)  \notag \\ &
+\frac{1}{4}\,c_{QtQb}^{(8)}\,\frac{m_b}{m_t}\!\left(
-\,C_A(18m_b^{2}+7m_t^{2})
+2C_F(36m_b^{2}+7m_t^{2})
\right), \notag \\
\left(\delta Z_{m_t,4q}^{(2,-1)}\right)_{\mathrm{BMHV}} = & \, \frac{8}{3}\,c_{QQ}^{(1)}\,m_t^{2}
+ \frac{16}{3}\,c_{tt}\,m_t^{2}
+ 3\,c_{Qb}^{(8)}\,m_b^{2}
+ 3\,c_{tb}^{(8)}\,m_b^{2} \notag \\ &
+ c_{QQ}^{(8)}\!\left(
3\,m_b^{2}
+ \frac{(9-4C_A+8C_F)\,m_t^{2}}{3}
\right)
- 24\,c_{Qt}^{(1)}\,m_t^{2} \notag\\ &
+ c_{Qt}^{(8)}\!\left(
3\,m_b^{2}
- 2(-3+2C_A+12C_F)\,m_t^{2}
\right) \notag\\ &
+ c_{QtQb}^{(1)}\!\left(
\frac{4(4+C_A)\,m_b^{3}}{m_t}
+ \frac{m_b m_t}{6}
+ \frac{2 m_t^{2}}{3}
\right) \notag\\ &
+ c_{QtQb}^{(8)}\!\left(
-\frac{2(1+2C_A-8C_F)\,m_b^{3}}{m_t}
- \frac{(C_A-2C_F)\,m_b m_t}{12} \right. \notag \\ & \left. 
- \frac{(C_A-2C_F)\,m_t^{2}}{3}
\right).
\end{align}
The results for the bottom quark mass is obtained by the replacement rule defined in eq.~\eqref{eq:mtmb-replacement}.
The renormalisation factors for the redundant class II operators in eq.~\eqref{eq:redundant-2-point} are shown in the electronic format in the notebook attached to the publication.

\section{Definition of evanescent operators}
\label{sec: Evanescent operators}
In this section, we show the sets of evanescent operators defining our schemes with a special focus on the ones that are generated at one-loop by the insertion of physical operators into Green's functions, but that also mix back in the physical sector by contributing to the quark self-energy. In this way, we have an overall single pole contribution from the evanescent operators to the two-loop renormalisation factors of the quarks. 

In Table~\ref{tab: fierz-evanescent} we show the Fierz-evanescent operators 
%(see Section~\ref{sec: X} for the discussion) 
where only $\mathcal{E}_{QtQb^{(R)}}^F$ are generated at one loop and which contribute to the quark self-energy in the NDR-scheme. 
In Table~\ref{tab:NDR-evanescent} we show the evanescent operators in the NDR-scheme which contribute to our results. In Tables~\ref{tab:penguin-evanescent} and \ref{tab:four-quark-evanescent} we show the two- and four-quark relevant evanescent operators in the BMHV-scheme, respectively. 
\small
\renewcommand{\arraystretch}{2.2}
\begin{table}[h]
\centering
\begin{tabular}{|c|}
\hline
\small Set of Fierz-evanescent operators $\mathcal{E}^F$ \\
\hline \hline
$\mathcal{E}_{tt^{(8)}}^F = (\tbar \gamma^\mu T^A t)(\tbar \gamma_\mu T^A t) - \left(\dfrac{1}{2} - \dfrac{1}{2 N} \right) \cO_{tt} \ \ \ (t \rightarrow b)$   \\

$\mathcal{E}_{tb^{(1)}}^F = (\tbar \gamma^\mu b)(\bbar \gamma_\mu t)- 2 \cO_{tb}^{(8)}-\dfrac{1}{N} \cO_{tb}^{(1)}$   \\

$\mathcal{E}_{tb^{(8)}}^F = (\tbar \gamma^\mu T^A b)(\bbar \gamma_\mu T^A t) + \dfrac{1}{N} \cO_{tb}^{(8)}-\dfrac{N^2-1}{2 N^2} \cO_{tb}^{(1)}$   \\

$\mathcal{E}_{QQ^{(3,1)}}^F = \dfrac{1}{2}(\bar Q \gamma^\mu \tau^I Q)(\qbar \gamma_\mu \tau^I Q) -\left( \dfrac{2}{N}-1\right)\cO_{QQ}^{(1)}- 4 \cO_{QQ}^{(8)}$ \\

$\mathcal{E}_{QQ^{(3,8)}}^F = \dfrac{1}{2}(\bar Q \gamma^\mu T^A \tau^I Q)(\qbar \gamma_\mu T^A \tau^I Q) - \left(\dfrac{N^2-1}{N^2} \right) \cO_{QQ}^{(1)} + \left(\dfrac{N+2}{N} \right) \cO_{QQ}^{(8)}$  \\

$\mathcal{E}_{Qt^{S,(1)}}^F = (\qbar t)(\tbar Q) + \cO_{Qt}^{(8)} + \dfrac{1}{2 N} \cO_{Qt}^{(1)} \ \ \ (t \rightarrow b)$ \\

$\mathcal{E}_{Qt^{S,(8)}}^F = (\qbar T^A t)(\tbar T^A Q) - \dfrac{1}{2 N} \cO_{Qt}^{(8)} + \dfrac{N^2-1}{4 N^2} \cO_{Qt}^{(1)} \ \ \ (t \rightarrow b)$ \\ [1.3ex]\hline \hline 

$\mathcal{E}_{QtQb^{(1)}}^F = \left((\qbar^I \sigma^{\mu\nu} t )\epsilon_{IJ}(\qbar^J\sigma_{\mu\nu} b )+\mathrm{h.c.}\right) +4\left(1-\dfrac{2}{N}\right)\mathcal{O}_{QtQb}^{(1)}-16\,\mathcal{O}_{QtQb}^{(8)}$\\ [1.3ex]

$\mathcal{E}_{QtQb^{(8)}}^F =\left(( \qbar^I \sigma^{\mu\nu} T^A t )\epsilon_{IJ}(\qbar^J \sigma_{\mu\nu} T^A b)+\mathrm{h.c.}\right)- 4\dfrac{N^2-1}{N^2} \mathcal{O}_{QtQb}^{(1)}+4\left(1+\dfrac{2}{N}\right)\mathcal{O}_{QtQb}^{(8)}$\\[1.3ex]
\hline
\end{tabular}
\caption{The set of Fierz-evanescent operators defined by eqs.~\eqref{eq: four-quark} and which do not include charge-conjugated fields. Only the second subset of operators, $\mathcal{E}_{QtQb^{(R)}}^F$, is generated at one-loop and contributes to the quark self-energy.}
\label{tab: fierz-evanescent}
\end{table}

\begin{table}[h]
\centering
\begin{tabular}{|c|}
\hline
\small Set of 4F $\left(\mathcal{E}_{\psi^4}\right)_{\mathrm{NDR}}$ evanescent operators in the NDR-scheme \\ [1.2ex]
\hline \hline
$\mathcal{E}_{LR^{(3)},t}^{(1)} =  \bar Q \gamma^\mu \gamma^\nu \gamma^\rho Q \, \bar t\gamma_\mu \gamma_\nu \gamma_\rho t - 4 (1+c_{\mathrm{ev}}\varepsilon)\mathcal{O}_{Qt}^{(1)}$\\ [1.3ex]
$\mathcal{E}_{LR^{(3)},t}^{(8)} =\bar Q \gamma^\mu \gamma^\nu \gamma^\rho T^a Q \, \bar t\gamma_\mu \gamma_\nu \gamma_\rho T^a t - 4 (1+c_{\mathrm{ev}}\varepsilon)\mathcal{O}_{Qt}^{(8)}$\\
$\mathcal{E}_{LR^{(3)},b}^{(1)} =  \bar Q \gamma^\mu \gamma^\nu \gamma^\rho Q \, \bar b\gamma_\mu \gamma_\nu \gamma_\rho b - 4 (1+c_{\mathrm{ev}}\varepsilon)\mathcal{O}_{Qb}^{(1)}$\\ [1.3ex]
$\mathcal{E}_{LR^{(3)},b}^{(8)} =\bar Q \gamma^\mu \gamma^\nu \gamma^\rho T^a Q \, \bar b\gamma_\mu \gamma_\nu \gamma_\rho T^a b - 4 (1+c_{\mathrm{ev}}\varepsilon)\mathcal{O}_{Qb}^{(8)}$\\[1.3ex]
\hline
\end{tabular}
\caption{The set of evanescent operators obtained from the evanescent structure in eq.~\eqref{eq:NDRevanescent} appearing in the NDR-scheme which contribute to the quark self-energy.}
\label{tab:NDR-evanescent}
\end{table}

\renewcommand{\arraystretch}{2.2}
\begin{table}[h!]
\centering
\begin{tabular}{|c|}
\hline
\small Set of 2F $\left(\mathcal{E}_{\psi^2G}\right)_{\mathrm{BMHV}}$ evanescent operators in the BMHV-scheme \\ [1.2ex]
\hline \hline
$\mathcal{E}_{t G}^{R} = \bar t_{R} i \hat \gamma^\mu \overline{\gamma}^\nu
 T^a t_{R} G_{\mu \nu}^a$ 
\\
$\mathcal{E}_{b G}^{R} = \bar b_{R} i \hat \gamma^\mu \overline{\gamma}^\nu
 T^a b_{R} G_{\mu \nu}^a$
\\
$\mathcal{E}_{t G}^{L} = \bar t_{L} i \hat \gamma^\mu \overline{\gamma}^\nu
 T^a t_{L} G_{\mu \nu}^a$ 
\\
$\mathcal{E}_{b G}^{L} = \bar b_{L} i \hat \gamma^\mu \overline{\gamma}^\nu
 T^a b_{L} G_{\mu \nu}^a$ 
\\

$\mathcal{E}_{t GD}^{LR} = \bar t_L \hat{\gamma}^\nu T^a t_R D^\mu G_{\mu \nu}^a + \text{h.c.} $  \\
$\mathcal{E}_{b GD}^{LR} = \bar b_L \hat{\gamma}^\nu T^a b_R D^\mu G_{\mu \nu}^a + \text{h.c.} $ \\ [1.3ex] \hline \hline
$\mathcal{E}_{tG}^{LR} = i \bar t_{L}  \hat \tau^{\mu \nu}  T^a t_{R} G_{\mu \nu}^a + \text{h.c.}$   \\
$\mathcal{E}_{bG}^{LR} = i \bar b_{L}  \hat \tau^{\mu \nu}  T^a b_{R} G_{\mu \nu}^a + \text{h.c.}$   \\[1.3ex] \hline
\end{tabular}
\caption{Evanescent operators generated at one loop for the renormalisation of the off-shell penguin diagram in the BMHV-scheme. The first set of operators is generated in both the four- and $D$-dimensional schemes for the gauge-fermion vertex. The second set is only generated in the latter. $t$ and $b$ are the physical top and bottom fields, and $\hat \tau^{\mu \nu} = \frac{1}{2} \comm{\hat \gamma^\mu}{\hat \gamma^\nu}$.}
\label{tab:penguin-evanescent}
\end{table}
\normalsize

\renewcommand{\arraystretch}{2.2}
\begin{table}[h!]
\centering
\begin{tabular}{|c|}
\hline
\small Set of 4F $\left(\mathcal{E}_{\psi^4}\right)_{\mathrm{BMHV}}$ evanescent operators in the BMHV-scheme \\ [1.2ex]
\hline \hline
$\mathcal{E}_{S,RR,tt}^{(1)} = \bar t_{L}  \hat \gamma^\alpha t_R \, \bar t_{L}  \hat \gamma_\alpha t_R +  \mathrm{h.c.} $ 
\\
$\mathcal{E}_{S,RR,tt}^{(8)} = \bar t_{L}  \hat \gamma^\alpha T^a t_R \, \bar t_{L}  \hat \gamma_\alpha  T^a t_R + \mathrm{h.c.} $
\\
$\mathcal{E}_{T,RR,tt}^{(1)} = \bar t_{L}  \hat\gamma^\alpha \bar \sigma^{\mu\nu}t_R \, \bar t_{L}  \hat\gamma_\alpha \bar \sigma_{\mu\nu}t_R +  \mathrm{h.c.} $ 
\\
$\mathcal{E}_{T,RR,tt}^{(8)} = \bar t_{L}  \hat\gamma^\alpha \bar \sigma^{\mu\nu} T^a t_R \, \bar t_{L}  \hat\gamma_\alpha\bar \sigma_{\mu\nu}   T^a t_R + \mathrm{h.c.} $ 
\\
$\mathcal{E}_{S,RR,bb}^{(1)} = \bar b_{L}  \hat \gamma^\alpha b_R \, \bar b_{L}  \hat \gamma_\alpha b_R +  \mathrm{h.c.} $ 
\\
$\mathcal{E}_{S,RR,bb}^{(8)} = \bar b_{L}  \hat \gamma^\alpha T^a b_R \, \bar b_{L}  \hat \gamma_\alpha  T^a b_R + \mathrm{h.c.} $
\\
$\mathcal{E}_{T,RR,bb}^{(1)} = \bar b_{L}  \hat\gamma^\alpha \bar \sigma^{\mu\nu}b_R \, \bar b_{L}  \hat\gamma_\alpha \bar \sigma_{\mu\nu}b_R +  \mathrm{h.c.} $ 
\\
$\mathcal{E}_{T,RR,bb}^{(8)} = \bar b_{L}  \hat\gamma^\alpha \bar \sigma^{\mu\nu} T^a b_R \, \bar b_{L}  \hat\gamma_\alpha\bar \sigma_{\mu\nu}   T^a b_R + \mathrm{h.c.} $ 
\\
$\mathcal{E}_{V,RL,tb}^{(1)} = \bar t_{L}  \hat\gamma^\alpha \bar \gamma^\mu b_L \, \bar b_{R} \hat\gamma_\alpha \bar \gamma_\mu t_R +  \mathrm{h.c.} $
\\
$\mathcal{E}_{V,RL,tb}^{(8)} = \bar t_{L}  \hat\gamma^\alpha \bar \gamma^\mu T^a b_L \, \bar b_{R} \hat\gamma_\alpha \bar \gamma_\mu T^a t_R +  \mathrm{h.c.} $
\\ [1.3ex] \hline \hline

$\mathcal{E}_{V,RL,tt}^{(1)} = \bar t_{L}   \hat\gamma^\alpha \hat\gamma^\beta \bar \gamma^\mu t_L \,\bar t_{R}  \hat\gamma_\alpha \hat\gamma_\beta \bar  \gamma_\mu t_R  $ 
\\
$\mathcal{E}_{V,RL,tt}^{(8)} =  \bar t_{L}  \hat\gamma^\alpha \hat\gamma^\beta \bar \gamma^\mu T^a t_L \,\bar t_{R} \hat\gamma_\alpha \hat\gamma_\beta \bar \gamma_\mu T^a t_R  $ \\
$\mathcal{E}_{V,RL,bb}^{(1)} = \bar b_{L}   \hat\gamma^\alpha \hat\gamma^\beta \bar \gamma^\mu b_L \,\bar b_{R}  \hat\gamma_\alpha \hat\gamma_\beta \bar  \gamma_\mu b_R  $ 
\\
$\mathcal{E}_{V,RL,bb}^{(8)} =  \bar b_{L}  \hat\gamma^\alpha \hat\gamma^\beta \bar \gamma^\mu T^a b_L \,\bar b_{R} \hat\gamma_\alpha \hat\gamma_\beta \bar \gamma_\mu T^a b_R  $ \\
$\mathcal{E}_{S,RL,tb}^{(1)} = \bar t_{L}   \hat\gamma^\alpha \hat\gamma^\beta  t_R \,\bar b_{L}  \hat\gamma_\alpha \hat\gamma_\beta b_R - \bar t_{L}   \hat\gamma^\alpha \hat\gamma^\beta  b_R \,\bar b_{L}  \hat\gamma_\alpha \hat\gamma_\beta t_R + \mathrm{h.c.}  $ 
\\
$\mathcal{E}_{S,RL,tb}^{(8)} = \bar t_{L}   \hat\gamma^\alpha \hat\gamma^\beta  T^a t_R \,\bar b_{L}  \hat\gamma_\alpha \hat\gamma_\beta T^a b_R - \bar t_{L}   \hat\gamma^\alpha \hat\gamma^\beta  T^a b_R \,\bar b_{L}  \hat\gamma_\alpha \hat\gamma_\beta T^a t_R + \mathrm{h.c.}  $ \\[1.3ex]  \hline
\end{tabular}
\caption{Evanescent operators generated at one-loop for the renormalisation of the four-quark vertex which contribute to the quark self-energy. The first set of operators is generated in both the four- and $D$-dimensional schemes for the gauge-fermion vertex. The second set is only generated in the latter. $t$ and $b$ are the physical top and bottom fields.}
\label{tab:four-quark-evanescent}
\end{table}
\normalsize 
\newpage
\section{Definitions for the results in electronic format}
In the notebook in the ancilliary files we present the full set of expressions that are the results of the paper. The results are shown in both the $4-$ and $5-$flavour schemes. Specifically, in the former we consider all the operators of eq.~\eqref{eq: four-quark} and both the top- and  bottom-quark to be massive. In the latter, the bottom quark is assumed to be massless, and the operators $\mathcal{O}_{QtQb}^{(R)}$ are excluded. The user can switch between the schemes with the setting \texttt{\$FlavourScheme} which takes the values \texttt{4,5}. 

We now describe how we presented the results of the paper. Let us start from the results obtained in the \MSbar\ scheme. All the results are collected in the following expressions:
\begin{itemize}
    \item \texttt{Zq2[type, scheme, N\_loops]} : Two-point renormalisation functions up to two loops proportional to the physical Wilson coefficients.
    \item \texttt{Zq2ev[type, scheme]} : Two-point renormalisation functions at one-loop proportional to the evanescent Wilson coefficients.
    \item \texttt{Z4Q[physical\_coeff]} : One-loop counterterms for the physical Wilson coefficients.
    \item \texttt{ZevNDR[evanescent\_coeff\_NDR]} : One-loop counterterms for the evanescent Wilson coefficients in the NDR scheme.
    \item \texttt{ZevBMHV[evanescent\_coeff\_BMHV]} : One-loop counterterms for the evanescent Wilson coefficients in the BMHV scheme.
    \item \texttt{$\gamma$[m$_\psi$, scheme, N\_loops]} : Anomalous dimension for the quark masses up to two loops.
\end{itemize}
The arguments are
\begin{itemize}
    \item \texttt{type} = \texttt{"m}$_{\texttt{t}}$\texttt{"},
\texttt{"t}$_{\texttt{L}}$\texttt{"},
\texttt{"t}$_{\texttt{R}}$\texttt{"},
\texttt{"}\(\delta^{3}\)\texttt{t}$_{\texttt{L}}$\texttt{"},
\texttt{"}\(\delta^{3}\)\texttt{t}$_{\texttt{R}}$\texttt{"},
\texttt{"}\(\delta^{2}\)\texttt{t}$_{\texttt{LR}}$\texttt{"},
\texttt{"m}$_{\texttt{b}}$\texttt{"},
\texttt{"b}$_{\texttt{L}}$\texttt{"},
\texttt{"b}$_{\texttt{R}}$\texttt{"},
\texttt{"}\(\delta^{3}\)\texttt{b}$_{\texttt{L}}$\texttt{"},
\texttt{"}\(\delta^{3}\)\texttt{b}$_{\texttt{R}}$\texttt{"},
\texttt{"}\(\delta^{2}\)\texttt{b}$_{\texttt{LR}}$\texttt{"},
    \item \texttt{m$_\psi$} = \texttt{"m}$_{\texttt{t}}$\texttt{"}, \texttt{"m}$_{\texttt{b}}$\texttt{"},
    \item \texttt{scheme} = \texttt{"NDR"}, \texttt{"BMHV"},
    \item \texttt{N\_loops} = \texttt{"1"}, \texttt{"2"},
    \item \texttt{physical\_coeff} =\texttt{"c}$_{\texttt{Qt}}^{(1)}$\texttt{"},
\texttt{"c}$_{\texttt{Qt}}^{\texttt{(8)}}$\texttt{"},
\texttt{"c}$_{\texttt{QQ}}^{\texttt{(1)}}$\texttt{"},
\texttt{"c}$_{\texttt{QQ}}^{\texttt{(8)}}$\texttt{"},
\texttt{"c}$_{\texttt{Qb}}^{\texttt{(1)}}$\texttt{"},
\texttt{"c}$_{\texttt{Qb}}^{\texttt{(8)}}$\texttt{"},
\texttt{"c}$_{\texttt{tt}}$\texttt{"},
\texttt{"c}$_{\texttt{bb}}$\texttt{"},
\texttt{"c}$_{\texttt{tb}}^{\texttt{(1)}}$\texttt{"},
\texttt{"c}$_{\texttt{tb}}^{\texttt{(8)}}$\texttt{"},
\texttt{"c}$_{\texttt{QtQb}}^{\texttt{(1)}}$\texttt{"},
\texttt{"c}$_{\texttt{QtQb}}^{\texttt{(8)}}$\texttt{"},
\item \texttt{evanescent\_coeff\_NDR} = \texttt{"k}$_{\texttt{QtQb}^{\texttt{(1)}}}^{\texttt{F}}$\texttt{"},
\texttt{"k}$_{\texttt{QtQb}^{\texttt{(8)}}}^{\texttt{F}}$\texttt{"},
\texttt{"k}$_{\texttt{LR}^{\texttt{(3)}}\texttt{,t}}^{\texttt{(1)}}$\texttt{"},
\texttt{"k}$_{\texttt{LR}^{\texttt{(3)}}\texttt{,t}}^{\texttt{(8)}}$\texttt{"},
\texttt{"k}$_{\texttt{LR}^{\texttt{(3)}}\texttt{,b}}^{\texttt{(1)}}$\texttt{"},
\texttt{"k}$_{\texttt{LR}^{\texttt{(3)}}\texttt{,b}}^{\texttt{(8)}}$\texttt{"},
\item \texttt{evanescent\_coeff\_BMHV} =  \texttt{"k}$_{\texttt{S,RR,tt}}^{\texttt{(1)}}$\texttt{"},
\texttt{"k}$_{\texttt{S,RR,tt}}^{\texttt{(8)}}$\texttt{"},
\texttt{"k}$_{\texttt{T,RR,tt}}^{\texttt{(1)}}$\texttt{"},
\texttt{"k}$_{\texttt{T,RR,tt}}^{\texttt{(8)}}$\texttt{"},
\texttt{"k}$_{\texttt{S,RR,bb}}^{\texttt{(1)}}$\texttt{"},
\texttt{"k}$_{\texttt{S,RR,bb}}^{\texttt{(8)}}$\texttt{"},
\texttt{"k}$_{\texttt{T,RR,bb}}^{\texttt{(1)}}$\texttt{"},
\texttt{"k}$_{\texttt{T,RR,bb}}^{\texttt{(8)}}$\texttt{"},
\texttt{"k}$_{\texttt{V,RL,tb}}^{\texttt{(1)}}$\texttt{"},
\texttt{"k}$_{\texttt{V,RL,tb}}^{\texttt{(8)}}$\texttt{"},
\texttt{"k}$_{\texttt{V,RL,tt}}^{\texttt{(1)}}$\texttt{"},
\texttt{"k}$_{\texttt{V,RL,tt}}^{\texttt{(8)}}$\texttt{"},
\texttt{"k}$_{\texttt{V,RL,bb}}^{\texttt{(1)}}$\texttt{"},
\texttt{"k}$_{\texttt{V,RL,bb}}^{\texttt{(8)}}$\texttt{"},
\texttt{"k}$_{\texttt{S,RL,tb}}^{\texttt{(1)}}$\texttt{"},
\texttt{"k}$_{\texttt{S,RL,tb}}^{\texttt{(8)}}$\texttt{"},
\texttt{"k}$_{\texttt{tG}}^{\texttt{R}}$\texttt{"},
\texttt{"k}$_{\texttt{tG}}^{\texttt{L}}$\texttt{"},
\texttt{"k}$_{\texttt{tGD}}^{\texttt{LR}}$\texttt{"},
\texttt{"k}$_{\texttt{tG}}^{\texttt{LR}}$\texttt{"},
\texttt{"k}$_{\texttt{bG}}^{\texttt{R}}$\texttt{"},
\texttt{"k}$_{\texttt{bG}}^{\texttt{L}}$\texttt{"},
\texttt{"k}$_{\texttt{bGD}}^{\texttt{LR}}$\texttt{"},
\texttt{"k}$_{\texttt{bG}}^{\texttt{LR}}$\texttt{"},
\texttt{"k}$_{\texttt{QtQb}^{\texttt{(1)}}}^{\texttt{F}}$\texttt{"},
\texttt{"k}$_{\texttt{QtQb}^{\texttt{(8)}}}^{\texttt{F}}$\texttt{"}.
\end{itemize}
We provide some examples: the counterterm for the mass of the top-quark in the BMHV-scheme at two loops is obtained via
\begin{equation}
    \left(\delta Z_{m_t, 4q}^{(2)}\right)_{\mathrm{BMHV}} = \texttt{Zq2["m$_{\texttt{t}}$", "BMHV", "2"]}\,.
\end{equation}
The running of the bottom quark mass (only in the 4FS) at two loops is obtained via
\begin{equation}
    \left(\gamma_{m_b, 4q}^{(2)}\right)_{\mathrm{BMHV}} = \texttt{$\gamma$["m$_{\texttt{b}}$", "BMHV", "2"]}\,,
\end{equation}
or for the evanescent contribution to the mass of the top counterterm in the NDR-scheme
\begin{equation}
    \delta m_{t, ev}^{\mathrm{NDR}} = \texttt{Zq2ev["m$_{\texttt{t}}$", "NDR"]}.
\end{equation}
Regarding the results in the on-shell scheme we have:
\begin{itemize}
    \item \texttt{Zq2OSb[type\_OS, scheme, N\_loops]} : Two-point on-shell renormalisation functions up to two loops in terms of the bare physical Wilson coefficients.
    \item \texttt{Zq2OSr[type\_OS, scheme, N\_loops]} : Two-point on-shell renormalisation functions up to two loops in terms of the \MSbar-renormalised physical Wilson coefficients.
    \item \texttt{Zq2evOS[type\_OS, scheme]} : Two-point on-shell renormalisation functions at one-loop proportional to the evanescent Wilson coefficients,
    \item \texttt{MSOSRelation[m$_\psi$, scheme, N\_loops]} : Relation between the on-shell and the \MSbar \ renormalised mass up to two loops,
\end{itemize}
where \texttt{type\_OS} = \texttt{"M}$_{\texttt{t}}$\texttt{"},
\texttt{"t}$_{\texttt{L}}$\texttt{"},
\texttt{"t}$_{\texttt{R}}$\texttt{"},
\texttt{"M}$_{\texttt{b}}$\texttt{"},
\texttt{"b}$_{\texttt{L}}$\texttt{"},
\texttt{"b}$_{\texttt{R}}$\texttt{"}.
As an example, the two-loop on-shell renormalisation functions for the top-quark mass and the top-quark left-handed field in terms of renormalised Wilson coefficients are obtained via
\begin{equation}
\begin{split}
     \left(\delta Z_{M_t,4q}^{(2)}\right)_{\mathrm{BMHV}} &= \texttt{Zq2OSr[\texttt{"M}$_{\texttt{t}}$\texttt{"}, "BMHV", "2"]}\,, \\ \left(\delta Z_{t,4q}^{L,(2)} \right)_{\mathrm{BMHV}}&=\texttt{Zq2OSr[\texttt{"t}$_{\texttt{L}}$\texttt{"}, "BMHV", "2"]}\,.
\end{split}
\end{equation}
\bibliographystyle{JHEP.bst}
\bibliography{main.bib}

\end{document}